\RequirePackage{ifpdf}
\documentclass[12pt]{JHEP3}
\usepackage{graphicx}
\usepackage{cite}
 \usepackage{ae}
 \usepackage{aecompl}
\usepackage{amsmath,epsfig}
\usepackage{amssymb,amsfonts}
\usepackage{latexsym}
\usepackage{epsfig}
\usepackage[caption=false]{subfig}
\usepackage{empheq}

\relax
\renewcommand{\theequation}{\arabic{section}.\arabic{equation}}

\def\D{\mathrm{d}}
\def\be{\begin{equation}}
\def\ee{\end{equation}}

\newcommand{\<}{\langle}
\renewcommand{\>}{\rangle}
\newcommand{\de}{\partial}
\newcommand{\bear}{\begin{eqnarray}}
\newcommand{\bea}{\begin{eqnarray}}
\newcommand{\eear}{\end{eqnarray}}
\newcommand{\eea}{\end{eqnarray}}
\newbox\pippobox

\def\II{\relax{\rm I\kern-.18em I}}

\title{Dressing the Electron Star in a Holographic Superconductor}

\author{Francesco~Nitti$^{a}$, Giuseppe~Policastro$^b$, Thomas~Vanel$^c$
~\\
$^a$ APC, Universit\'e Paris 7, CNRS/IN2P3, CEA/IRFU, Obs. de Paris, Sorbonne Paris Cit\'e, B\^atiment Condorcet, F-75205, Paris Cedex 13, France (UMR du CNRS 7164)
~\\
$^b$ Laboratoire de Physique Th\'eorique, Ecole Normale Sup\'erieure, 24 rue Lhomond, 75231 Paris Cedex 05, France (UMR du CNRS 8549)
~\\
$^c$ Laboratoire de Physique Th\'eorique et Hautes Energies, Universit\'e Pierre et Marie
Curie, 4 place Jussieu, 75252 Paris Cedex 05, France (UMR du CNRS 7589)
~\\
}

\preprint{LPTENS 13/17}

\abstract{We construct new asymptotically $AdS_4$ solutions dual to 2+1 CFTs at finite density and zero temperature  by combining the ingredients of the electron star and the holographic superconductor. The solutions, which we call {\em compact electron stars}, contain both a fermionic fluid and charged scalar hair in the bulk. We show that the new solutions are thermodynamically favoured in the region of parameter space where they exist. Along the boundary of this region, we find evidence for a continuous phase transition between the holographic superconductor and the compact star solution.}

\begin{document}

\maketitle 

\section{Introduction} \label{intro}
In the last years, the holographic approach has been fruitfully used to construct new examples of ground states of interacting systems at finite density, reaching beyond the validity of the traditional tools. 
 
The conventional field theoretic methods, that are reliable at weak coupling, have been remarkably successful in accounting for the properties of many 
classes of fermionic systems.  The properties of the ground state are characterized in terms of the symmetry breaking pattern. If no symmetry is broken, the only ground state 
is the Fermi liquid, which is adiabatically connected to the ground state of free fermions. Landau's theory of normal Fermi liquids is sufficient to describe the properties of most metals. 
With a broken (global/gauge) $U(1)$ charge the system is a superfluid/superconductor respectively and the BCS theory gives an adequate description. Other symmetries may also be broken (e.g. translational and rotational invariance) possibly also in combination with the internal symmetries and this accounts for the large variety of materials that we observe in the world. 

However  there are many other systems that do not fall into this classification and for which the weak coupling techniques are insufficient (Quantum Hall effect, high-Tc superconductors, heavy fermions materials, etc.). 
The non-triviality of the ground state may be encoded in subtler properties that are not revealed simply by the expectation value of local operators (entanglement, topological order). In this context, the holographic method is proving to be very valuable in opening a new window on these fascinating questions and offering constructible and solvable examples of non-trivial ground states \cite{Hartnoll:2009sz, Herzog:2009xv, McGreevy:2009xe}.
A particularly important question that can be addressed is the existence of other vacua with unbroken symmetries that are not connected to the Fermi liquid. There are several reasons to think that these should exist; an experimental one comes from the phase diagram of cuprate superconductors. Outside the region of superconductivity, there are two metallic phases, at large doping and small doping, where the system has a Fermi surface that has been observed with photoemission experiments; it was found that the Fermi surface has a different volume on the two sides, suggesting that a dramatic change has happened in passing close to the critical point. The Luttinger theorem that relates the volume of the Fermi surface to the total charge must be violated in one of the two phases, so that only a fraction of the charge carriers contributes to the Fermi surface. One explanation is that the charge degrees of freedom have become ``fractionalized", a process that may be thought as the inverse of confinement: the fundamental particles, i.e. the electrons, effectively break down and are 
replaced by quasiparticles that carry only a fraction of the charge. 

Quite remarkably, this is precisely what is observed also in holographic models. In the holographic setup, the simplest ground state at finite density is dual to the AdS-RN extremal black hole. The charge of the black hole 
is the same as the charge density at the boundary, but it is completely inaccessible to gauge invariant observables and does not create a Fermi surface, so one may think that this system is totally fractionalized. 
It was found however in \cite{Hartnoll:2009ns} that this vacuum is actually unstable to the creation of matter in the bulk of the spacetime, generating the ``electron star" solution \cite{Hartnoll:2010gu}. In this system, the charge is sourced by 
matter fields in the bulk that are dual to gauge-invariant operators, and they exhibit a Fermi surface. Turning on relevant deformations it is possible to find other phases that are only partially fractionalized, so that the charge is partly hidden behind the horizon and partly visible in the bulk fermions, with a phase transition that can be continuous or first order depending on the parameters \cite{Hartnoll:2011pp} (see also \cite{Adam:2012mw} for an alternative route). 

The spontaneous breaking of an internal $U(1)$ symmetry can also be modeled in holography using a different instability of charged black hole, with respect to forming a ``scalar hair". 
The reason for the instability is that via the coupling to the gauge field, the scalar acquires an effective mass that in the near-horizon region goes below the Breitenlohner-Freedman bound. 
The endpoint of the instability is the geometry found in \cite{Hartnoll:2008vx,Hartnoll:2008kx,Gubser:2008pf,Gubser:2009cg,Horowitz:2009ij} that has been named ``holographic superconductor". Strictly speaking it is an abuse of language since in the boundary theory the broken $U(1)$ is not gauged, 
but one can imagine gauging it and computing the effects in perturbation theory in the gauge coupling; indeed one finds that the system behaves as a superconductor, with zero dc resistivity and a gap in the 
spectral function of the current at low frequency.  
The charged scalar field in the bulk is dual to an operator of the field theory whose condensation breaks the $U(1)$ and should be thought of as the strong-coupling analogue of a Cooper pair. 

There are then two unrelated instabilities from which a charged black hole can suffer; it is natural to ask if there can be a competition between the two. If both fermionic matter and charged scalars 
are present in the model, one can distribute the charge between the two and have a richer family of solutions with an interesting phase diagram. This is the problem we set up to investigate in the present paper. 
It is an instance of an ubiquitous phenomenon in interacting systems in which the structure of a phase diagram is determined by the competition between several possible instabilities, each coming typically with a 
different symmetry pattern. 

In the electron star geometry, the presence of charged matter in the bulk is responsible for a screening of the electric field, the photon becomes effectively massive and 
the IR geometry close to the horizon becomes a Lifshitz solution, with a dynamical scaling exponent that depends on the parameters of the model. This infrared part of the geometry represents a sector of 
low-energy excitations that are critical and interact with the charged matter. As a Lifshitz spacetime is singular, at present it is still unclear whether this can be considered as the true ground state of the system. 

As in \cite{Hartnoll:2010gu}, we will work in the analog of the Thomas-Fermi approximation in which the fermions in the bulk can be described as a classical charged ideal fluid. The approximation is valid when the fermions are at high enough density, and their Compton wavelength is smaller than the characteristic scale of the geometry. On the boundary side, this means that instead 
of a single Fermi surface, there is a large number of them, closely spaced, so that effectively one sees a continuum of zero-energy excitations \cite{Hartnoll:2010xj,Hartnoll:2011dm}. 
This generates some unphysical features that can be resolved by considering the quantization of the bulk fermions \cite{Sachdev:2011ze,Allais:2012ye,Allais:2013lha}, but the problem becomes technically much more challenging. 

Phases where a Fermi surface coexists with a superfluid have been considered
in \cite{Huijse:2011hp} in field theories that are relevant for condensed matter systems and
for gauge-gravity duality. In the holographic approach, a closely related system
to the one we study here was considered by \cite{Edalati:2011yv}, with the difference that they
have a neutral scalar in the bulk, so the phase transition is not to a superconductor but to
antiferromagnetic or nematic phases, or other 
transitions characterized by a neutral order parameter. Another related work \cite{Adam:2012mw}
considers also a competition  between fractionalization and superconductivity, but with a bulk
system consisting of a charged scalar and a neutral one (a dilaton). So in both cases, there is
only one field that can carry the charge away from the horizon and the possibility that we have
in mind cannot be realized. 
 
Our main results  is that we found  novel  solutions in which  a condensate for the scalar field coexists with an electron star (we call it a ``compact electron star,'' because the fermionic fluid does not extend all the way down to the horizon); we  find that this solution, when it exists,  has lower free energy than  the solutions where only the condensate or only the fluid are present. We chose to parametrize the solutions in terms of the mass of the fermions and the charge of the scalar, for different values of the scalar mass. We find that the compact star exists only in a certain region of this parameter space, and on the boundary of this region it merges with the holographic superconductor solution. Along this boundary, there is a continuous phase transition between the superconductor phase  and the compact star, across which the free energy changes smoothly. 
Our results suggest also that, for a certain range of values of the scalar mass, there could be another phase boundary between the electron star and the compact star, around the point of onset of instability for the scalar around the black hole, but we could not determine its presence with certainty with the limits of our numerical accuracy.  

We have to point out that while the transfer of charge from one subsystem to another is perfectly natural and evident from the bulk point of view, it is not easy to see or to interpret it in the boundary theory; 
the data that one reads off from the asymptotic values of the fields at the boundary are only related to the total charge of the system and the vev of the scalar condensate. In order to see the effect on the Fermi 
surface one should consider the fluctuations of the fermions on the background. We will leave this for future work, as well as considering the effects of finite temperature. 

The plan of the paper is as follows: in section \ref{reviewsec} we review the solutions already known - the planar AdS-RN black hole, the electron star and the holographic superconductor; in section \ref{recipe} we define the 
model that we are solving (Einstein-Maxwell + charged fluid + charged massive scalar) and describe the procedure to find the solutions, the asymptotic boundary conditions and the parameter space. In section \ref{phase} we 
give our results for the phase diagram as a function of the parameters. 
We conclude in section \ref{discussion}  indicating open problems and directions for further work. 

In the appendices,  \ref{Fieldeq} contains details about the action and the equations of motion of the model, 
\ref{AppOnshell} contains the derivation of the thermodynamical quantities and the verification that the first law is satisfied, \ref{probe} contains the study of the charged scalar as a probe of the electron star, which was a prerequisite to the fully backreacted system, and \ref{Lifshitz} presents some solutions in which the scalar field is dual to an irrelevant operator in the UV.  

While working on this paper we learned that another work strongly overlapping with ours was in
preparation \cite{Liu:2013yaa}.

\section{Review of the charged zero-temperature solutions} \label{reviewsec}
We will consider zero-temperature gravitational solutions which share
the following two features:
\begin{itemize}
\item They are asymptotically $AdS_4$
\item They have zero temperature and finite electric charge. 
\end{itemize}
There are different kinds of models in (3+1) dimensions which have appeared so far and 
allow these solutions, and they are described generically from an action  of the form
\begin{equation} \label{action}
S = \int \D^4x \sqrt{-g}\left[\frac{1}{2\kappa^2}\left(R+\frac{6}{L^2}\right)-\frac{1}{4e^2}F_{ab}F^{ab} \right]+ S_{bdr} + S_{matter} \ ,
\end{equation}
where $\kappa$ is Newton's constant, $L$ is the asymptotic $AdS_4$  length, and $e$ the $U(1)$ coupling, and $S_{bdr}$ represents collectively the Gibbons-Hawking term and the boundary counterterms needed for holographic renormalization.  From now on we will set $L=1$. 

Without loss of generality, we will take the metric  $\D s^2$ and Maxwell one-form $A$ to be
\be\label{sol}
\D s^2 = -f(r)\D t^2 + g(r)\D r^2 + \frac{1}{r^2} \left(\D x^2 + \D y^2\right), \quad
A = {e\over \kappa} h(r) \D t \ ,
\ee
in which $r=0$ is the AdS boundary.  
Any homogeneous and spatially isotropic solution can be brought to this form by suitable diffeomorphisms and gauge transformations. 

The simplest model  is the pure  Einstein-Maxwell system, 
which admits charged extremal black holes; adding charged
matter leads to the Electron Star \cite{Hartnoll:2010gu} (matter is a charged fermionic
fluid) and the Holographic Superconductors \cite{Herzog:2009xv} (matter is a charged scalar  field with $m_s^2 <0$). Below, we review these three models. 
Introducing more scalar fields can add features to these solutions, like partial fractionalization, as seen for example in \cite{Edalati:2011yv,Adam:2012mw,Hartnoll:2011pp}.

\subsection{Extremal  Black Hole}
Extremal Reissner-Nordstrom-AdS (ERN for short) black holes are  solutions of the pure Einstein-Maxwell theory with $S_{matter}=0$. They are characterized by a single parameter, i.e. their charge. The metric and 
electric potential are given by
\be\label{bh1}
f(r) = {1\over r^2}\left(1 - M r^3 +  {Q^2 \over 2} r^4\right) \ , \quad g(r) = {1\over r^4 f(r)} \ , \quad h(r) = \mu - Q r \ .
\ee   
The outside geometry extends from the boundary $r=0$ to the outer horizon $r_+$ where $f = h =0$. 
The parameters $M$ and $Q$ are the mass and charge density of the black hole, and  $\mu$ can be thought of as  a chemical potential. These quantities correspond to the energy, charge density and chemical potential of the dual theory. They are related to the outer horizon radius by
\be\label{bh2}
M = {4\over r_+^3} \ , \qquad Q = {\sqrt{6}\over r_+^2} \ ,  \qquad \mu= {\sqrt{6}\over r_+} \  .
\ee
Close to the horizon $r_+$, $f(r)$ vanishes quadratically, 
\be\label{bh3}
f(r) \sim {6\over r_+^4}(r_+ - r)^2 \ , 
\ee
therefore the temperature of this solutions is zero.
With the change of variables $\rho = (r_+^2/6) (r_+-r)^{-1}$, the near-horizon geometry becomes
\be\label{bh4}
\D s^2 \sim  {1\over 6\rho^2}\left(-\D t^2 + \D \rho^2\right) + {1\over r_+^2}\left(\D x^2 + \D y^2\right),  \quad h(r) \sim {\sqrt{6} \over r_+^2}(r_+-r) \ ,
\ee 
that is, $AdS_2\times R^2$, with the $AdS_2$ radius given by $1/\sqrt{6}$. 

\subsection{Electron Star}

The electron star (ES) is obtained by coupling the Einstein-Maxwell system to bulk charged fermions, of mass $m_f$ and charge $q_f$,  in the approximation that the fermionic degrees of freedom can be described by a degenerate Fermi gas. Although one can give a Lagrangian description of the fluid, for our purposes it would be sufficient to introduce it in Einstein's equations via its equilibrium stress-energy tensor, characterized by energy density $\rho(r)$ and pressure $p(r)$, and its $U(1)$ charge density $\sigma(r)$. These quantities are assumed to satisfy the chemical equilibrium equation of state of a Fermi gas, with a density of states given by
\be\label{star1}
n(E) 
\left\{\begin{array}{lr}
\propto E\sqrt{E^2-m_f^2} & ~~ E > m_f \ , \\
 =  0 & ~~ 0<E<m_f  \ .
 \end{array}\right.
\ee
The energy density $\rho$, charge density $\sigma$, and pressure $p$, are then given by
\be\label{star2}
\rho(r) = \beta \int_{m_f}^{\mu_l(r)} s\, n(s)\, \D s \ ,  \quad \sigma(r) = \beta \int_{m_f}^{\mu_l(r)} n(s)\, \D s \ , \quad  -p(r) = \rho(r) -\mu_l(r) \sigma(r) \ ,
\ee
where $\mu_l(r)$ is the {\em local} chemical potential in the bulk, given by
\be\label{star3}
\mu_l(r) = {h\over \sqrt{f}} \ ,
\ee 
and $\beta$ is a constant which together with $m_f$ is one of the parameters of the model. The quantities 
appearing in eq. (\ref{star1}-\ref{star3}) are suitable dimensionless combinations obtained by rescaling by appropriate powers of $e L/\kappa$. See \cite{Hartnoll:2010gu} and Appendix \ref{Fieldeq} for details.  
The definition (\ref{star3}) of the local chemical potential  can be obtained by writing the
coupling between the charge density and  the electromagnetic field in covariant form,
$\mathcal{L}_{coupling} = \sqrt{g} A_a u^a \sigma = \sqrt{g} \mu_l(r) \sigma(r)$,   and then
specifying to the fluid rest frame  with $u_t = \sqrt{f(r)}$ and $A_t = h(r)$. 

The solutions of the Einstein-Maxwell system coupled to the fermion fluid, characterized by non-zero fluid densities, exist only for $0\leq m_f < 1$, and are called Electron Stars \cite{Hartnoll:2010gu}; they have two regions separated by the star boundary $r_s$, where $\mu_l(r_s)= m_f$, and where the fluid density vanishes
\begin{itemize}
\item {\bf Inner region (IR):} The fluid density is non-zero  only in the region $r>r_s$, in which $\mu_l(r)> m_f$. The inner region has  a non-zero charge density, and  extends to $r \to \infty$ where the geometry is asymptotically Lifshitz, 
\be\label{star4}
f(r) \sim {1\over r^{2z}}, \qquad g(r)\sim {g_{\infty}\over r^2}, \qquad h(r) \sim {h_{\infty} \over r^{z}}  \ ,
\ee
in which the charge, energy density and chemical potential become constants, and the scaling exponent $z$ is fixed by the parameters of the model and is related to the asymptotic local chemical potential by
\be\label{star5}
h_{\infty} = \sqrt{z-1\over z} \ .
\ee
\item {\bf Outer region (UV):} Outside the star boundary, i.e. for $0< r< r_s$, the solution is the RN extremal black hole described in the previous section, i.e. eq. (\ref{bh1}) with charge equal to the total charge of the star, 
\be\label{star6}
Q = \int_{r_s}^{\infty} \D r\, {\sqrt{g} \over r^2}\sigma(r) \ .
\ee
The boundary chemical potential $\mu$ is fixed by ensuring the continuity of the solution at $r_s$. Notice that this is not the same as the {\em bulk} chemical potential $h/\sqrt{f}$, which vanishes as $r$ close to the boundary.

The electron star solution exists only in the range $0\leq m_f< 1$. As $m_f \to 1^{-}$ the Lifshitz exponent $z\to +\infty$. It can be shown (see e.g. \cite{Gouteraux:2012yr}) that in this limit the metric develops an   $AdS_2\times R^2$ horizon. Therefore,  as $m_f \to 1$ the compact star solution merges onto the extremal black hole solution. 
\end{itemize}
\subsection{Holographic Superconductor}
The holographic superconductor (SC) is obtained as a solution of the Einstein-Maxwell theory coupled to a charged scalar field, with action\footnote{Our conventions are adapted to match  the electron star papers conventions \cite{Hartnoll:2010gu}. In order to recover the holographic superconductor conventions of e.g.~\cite{Hartnoll:2008kx} and \cite{Horowitz:2009ij} one must
rescale $A_a\rightarrow(1/\sqrt{2})A_a$ and $q\rightarrow \sqrt{2}\, q$.}
\be\label{sc1}
S_{scalar} = -{1\over 2}\int \D^4 x \sqrt{-g}\left[ |\de_a \psi - iq A_a \psi|^2 + m_s^2
|\psi|^2 \right] \ ,
\ee
with charge $q$ and  mass squared $m_s^2$ negative, and above the $AdS_4$ BF bound , $-9/4 < m_s^2 < 0$.
The theory still admits the extremal charged black holes of Section 2.1, with trivial scalar. However if
the effective mass of the scalar is lower than the {\em infrared} $AdS_2$ BF bound, one expects the 
scalar to condense \cite{Hartnoll:2008kx}. This condition corresponds to
\be\label{sc2}
{m_s^2 - q^2 \over 6} < -1/4 \ .
\ee
The solution is again of the form (\ref{sol}), with the addition of a non-trivial scalar field profile, 
$\psi = \psi(r)$,   
which  breaks the $U(1)$ symmetry. These solutions are dual to a superfluid (or superconducting) phase of the boundary theory. They were first found at finite temperature, where it was shown that below a critical temperature $T_c$ they are favored with respect to the black hole solutions. 

The zero-temperature limit of the superconductor was discussed in \cite{Horowitz:2009ij}.
\begin{itemize}
\item {\bf In the UV,} as $r\to 0$ the solution approaches  $AdS_4$  as in (\ref{bh1}), with in addition a scalar field given by
\be\label{sc4}
\psi(r) \sim \psi_+ r^{\Delta_+} \ , \qquad \Delta_+ = {3\over 2} + {3\over 2}\sqrt{1 + {4m^2_s\over 9}} \ .
\ee
In this formula, $\psi_+$ is (proportional to) the vacuum expectation value  of the charged boundary operator $O$ dual to $\psi$,  and it determines the charge and chemical potential of the solution. 
\item {\bf In the IR,} as $r\to \infty$,  the solution is singular, and it behaves as follows:
\begin{equation}
\label{sc6}
\begin{aligned}
f(r) &\sim \frac{1}{r^2}  \ ,  &  g(r) \sim -\frac{3}{2m_s^2} \frac{1}{r^2 \log r} \ , 
 \\
h(r) &\sim  h_0 r^{\delta}\left(\log r\right)^{1/2} \ ,  & \psi(r) \sim 2\left(\log r \right)^{1/2} ~~~~   . 
\end{aligned}
\end{equation}
The parameter $h_0$ has to be fixed in the UV in order to set to zero the leading scalar field asymptotic solution $\psi_-\sim r^{\Delta_-}$. The exponent $\delta$ in (\ref{sc6}) is given by
\be\label{sc7}
\delta = \frac{1}{2} - \frac{1}{2}\left(1-\frac{24q^2}{m_s^2}\right)^{1/2} \ .
\ee
This is an asymptotic solution as $r\to \infty$ only if $\delta < -1$.  This means that this solution exists only if
\be \label{sc8}
q^2 > -m_s^2/3.
\ee 
\end{itemize}
\section{Dressing the Electron Star}\label{recipe}

The model that will be at the center of our study combines the elements described in the previous section. We consider now Einstein-Maxwell theory coupled to both a charged scalar, with mass square $m_s^2$ and charge $q$,  and a fluid made out of charged fermions of mass $m_f$. For convenience we set the charge of the fermions to one. 
The action is given by  eq. (\ref{action}), with the matter content given by the scalar field action (\ref{sc1}) plus the fluid. Rather than writing 
the fluid action, we insert its stress tensor and current density in Einstein and Maxwell equations. The relevant parameters are then (after we scale out appropriately the quantity $e/\kappa$)
\be
\textrm{scalar:} \: (q,  m_s), \qquad \textrm{fluid:} \: (m_f, \beta).
\ee
The details about the definition of the model and its parameters are given in Appendix \ref{Fieldeq}.
We restrict  to the case $0> m_s^2 > -9/4$ (i.e. $\psi$ is dual to a relevant operator on the field theory side) but for now we don't impose any condition on the other parameters. 
The sign of the scalar charge turns out to be irrelevant. 
Notice that there is no direct coupling between the fluid and the scalar field, nor are we including a scalar potential.

We will look for  fully backreacted solutions, with contributions to the metric from both the scalar field and the fluid.
The field equations are written explicitly in \ref{Fieldeq}.

First, one can consider the scalar field in the probe approximation on top of the electron star background. The analysis is carried out in Appendix \ref{probe}, where we conclude that for $m^2_s<0$,  if the mass is below a certain critical mass (corresponding to the Lifshitz BF bound) the scalar field develops a tachyonic instability and condenses. Above this bound, the solution that is normalizable in the UV will generically go to a non-normalizable one in the IR and so the condensation is not expected. In the same way, for fixed $m_s^2$, there is a critical value of $q$ above which the condensation occurs, just as for the black hole as discussed in 
eq. (\ref{sc2}).  However the effect of the backreaction can change this conclusion, as we will see. 

Next,  we turn to looking for solutions of the full system.
We search  for homogenous, time-independent  configurations  which preserve the rotational symmetry in the $(x,y)$ plane. The most general solution has the form
\begin{equation}
\label{rec1}
\begin{aligned}
\D s^2 &= -f(r)\D t^2 + g(r)\D r^2 + \frac{1}{r^2} \left(\D x^2 + \D y^2\right) \ , \\
A &= h(r) \D t, \qquad \psi= \psi(r), \quad \rho = \rho(r), \quad \sigma=\sigma(r).  
\end{aligned}
\end{equation}
Since the fluid densities are fixed in terms of the local chemical potential $\mu_l(r) = h/\sqrt{f}$ by (\ref{star2}),  they do not constitute independent unknowns, and the independent functions to solve for are $g(r),f(r),h(r),\psi(r)$.  In what follows, we denote  with a prime the derivatives with respect to the radial coordinate $r$.

First, consider the $r$-component of Maxwell equations:
\begin{equation}
\label{eq:phase-psi}
 q (\bar{\psi}\psi'-\psi\bar{\psi}') = 0 \,. 
\end{equation}
This implies that when the scalar condenses, the phase of $\psi$ is constant, and we can fix it to zero in the whole solution by a global $U(1)$ transformation. $\psi(r)$ can be now considered as a real field. 

The other equations of motion are
\begin{subequations}
\label{eq:system}
\begin{align}
\psi'' + \left(\frac{f'}{2f}-\frac{g'}{2g}-\frac{2}{r}\right)\psi' +
g\left(\frac{q^2h^2}{f}-m^2\right)\psi &= 0  \label{eq1}\\
h'' - \frac{1}{2}\left(\frac{f'}{f}+\frac{g'}{g}+\frac{4}{r}\right)h' -
g\left(\sqrt{f}\sigma+q^2h|\psi|^2\right) &= 0 \label{eq2} \\
g' + \left( \frac{5}{r}+\frac{r h'^2}{2f}+\frac{r}{2}|\psi'|^2 \right)g
+ \left[ \frac{r}{2}\left(\frac{q^2h^2}{f}+m^2\right)|\psi|^2 +r(\rho-3) \right] g^2 &=
0 \label{eq3} \\
f' + \left[ rg(p+3) - \frac{1}{r}  + \frac{1}{2}r|\psi'|^2 +
\frac{r}{2}g\left(\frac{q^2h^2}{f}-m^2\right)|\psi|^2 \right]f - \frac{1}{2}rh'^2 &= 0 \label{eq4}
\end{align}
\end{subequations}
where the  fluid functions $p(r)$, $\sigma(r)$ and $\rho(r)$ are given as functions of $\mu_l= h/\sqrt{f}$ by (\ref{star2}). We now discuss the universal UV asymptotic behavior of the solution, and more interestingly, the different possible IR solutions.

\subsection{UV asymptotics}
In the UV the metric should be asymptotically $AdS_4$: 
\be \label{uv1}
r\to 0:\qquad f(r)\sim g(r) \sim {1\over r^2}, \quad h(r) \sim \mu - Q r\,.   
\ee
Notice that, for any asymptotically $AdS_4$ solution, the local chemical  potential vanishes close to the boundary as 
\be\label{uv2}
 \mu_l \sim \mu r, \qquad r\to 0. 
\ee
 Therefore, for any finite $m_f$, the fluid will reach zero density at a finite radius $r_s$, defined by
\be\label{uv3}
m_f = \mu_l(r_s).  
\ee
Between $r_s$ and the UV boundary, only the scalar and gauge fields will be non-trivial, and  the UV asymptotic solution will  know nothing about the star, except that its charge will contribute to the electric flux. 

Close to the boundary, in general the scalar will be characterized by two real parameters $\psi_-$ and $\psi_+$, representing  the source and the expectation value of the corresponding field theory operator:
\be \label{uv4}
\psi(r) \sim \psi_- r^{(3-\Delta)} + \psi_+ r^{\Delta}, \qquad \Delta= {3\over 2}\left(1 + \sqrt{1 + {4m^2_s\over 9}}\right). 
\ee 
We take $m_s^2<0$, i.e.  $\Delta < 3 $, so that the dual  operator is relevant, and its condensation will drive the theory away from the UV fixed point. Also,  
we are interested in the case where the breaking of $U(1)$ is spontaneous, i.e. in our solution we impose the condition
\be\label{uv5}
\psi_- = 0. 
\ee
This will fix one of the non-trivial integration constants of the system. 

It is useful to have a handle of the asymptotics of the metric and gauge field beyond the leading order (\ref{uv1}), in particular to understand the corrections due to the condensate. At $\psi_- = 0$ we know the exact solution, i.e. the extremal Reissner-Nordstrom geometry (\ref{bh1})\footnote{Even if a star is present in the bulk, this solution is still exact in the exterior UV region.}. If we turn on a condensate, this will backreact on the  solution,  but since we are taking  the scalar field to be  a purely  normalizable mode  as $r\to 0$, we can compute the backreaction  perturbatively in eq. (\ref{eq:system}). The resulting deformed UV solution with the condensate turned on behaves, as $r\to 0$, as
\begin{equation}
\label{uvasym}
\begin{aligned}
 & f(r) = {1\over r^2}\left(1 - M r^3 +  {Q^2 \over 2} r^4 + \ldots\right) \ ,  \\
&g(r) = {1\over r^4 f(r)}\left(1- {\Delta \over 2}\psi_+^2 r^{2\Delta} +\ldots \right)\ , \\
& h(r) = \mu - Q r + \mu {q^2\psi_+^2 \over 2\Delta(2\Delta-1)}r^{2\Delta} +\ldots \ , \\ 
&\psi(r) = \psi_+ r^{\Delta} +\ldots \ , 
\end{aligned}
\end{equation}
where the dots denote terms which are subleading with respect to those we have included, and whose exact order is unimportant. 

Equations (\ref{uvasym}) solve the field equations for {\em arbitrary} values of $\mu,M,Q,\psi_+$.  Remarkably,  $f(r)$ receives no corrections to leading   order, and $h(r)$ and the  relation between $g(r)$ and $f(r)$ are corrected only by terms which are subleading with respect to {\em all} terms appearing in the exact  black hole solution (\ref{bh1}). This means that the   the condensate enters into the metric and gauge field  at subleading order in  $r\to 0$ with respect to the charge and mass parameters
of the solution.

\subsection{The Compact Star solution}

Given that the UV is fixed to be $AdS_4$, there are four branches of solutions. We denote the total charge by $Q$, and the scalar and fluid charges by $Q_s$ and $Q_f$ respectively. 

First of all, notice that all three possibilities reviewed in Section \ref{reviewsec} are also solutions of the general model: 
\begin{enumerate}
\item {\bf ERN:} The Extremal Reissner-Nordstrom black hole is a solution with $\psi = 0$ and $\sigma = \rho= 0$. The IR geometry is given in eq (\ref{bh4}) 
 The electric charge is all inside the horizon, $Q_s=Q_f=0$,  corresponding to a completely fractionalized phase. 
\item {\bf ES:} The Electron Star  is found by setting $\psi=0$ but matching the  RN  exterior to a solution with non-vanishing fluid density at the point 
where $h/\sqrt{f}=m_f$, and continuing it inward towards an asymptotically Lifshitz metric, eq. (\ref{star4}).  The charge is all in the fermion fluid, $Q=Q_f$, $Q_s=0$ .
\item {\bf SC:} The Holographic Superconductor is obtained by setting $\sigma=\rho = 0$ but allowing $\psi \neq 0$. The IR geometry is given in (\ref{sc6}).  In this case, the charge is all in the condensate, $Q=Q_s$, $Q_f=0$.  

\end{enumerate}

In addition to  these known branches, a new solution is now possible, with the same IR asymptotics as the SC solution: 

\begin{enumerate}
\setcounter{enumi}{3} 
\item  {\bf  \em Compact Star (CS):} This solution corresponds to  a  fluid density confined in a shell $r_2< r< r_1$, and with a non-zero scalar condensate. The IR geometry is the same as for the superconductor  solution, i.e. eq. (\ref{sc6}).
\end{enumerate}
To see how this solution can arise, notice that in the SC solution, the local chemical potential $h/\sqrt{f}$ vanishes both in the UV and in the IR, as can be seen from eq. (\ref{uv2}) and (\ref{sc6}). $\mu_l(r)$ has a maximum, and if its maximum value happens to be larger than $m_f$, there are two solutions $r_1$ and $r_2$  to the eq. (\ref{uv3}) defining the star boundary.  Thus, we can start from the IR with an SC solution. The local chemical potential will increase towards the UV and  then at the point $r_1$ we can match the solution with an interior with non-trivial density. The density will reach a maximum, then decrease again until it becomes zero at $r_2$. At this point, the solution is matched with a new SC solution up to the UV boundary. 

\begin{figure}[h!]
\centering
\subfloat[ ]{
\includegraphics[width=0.49\textwidth]{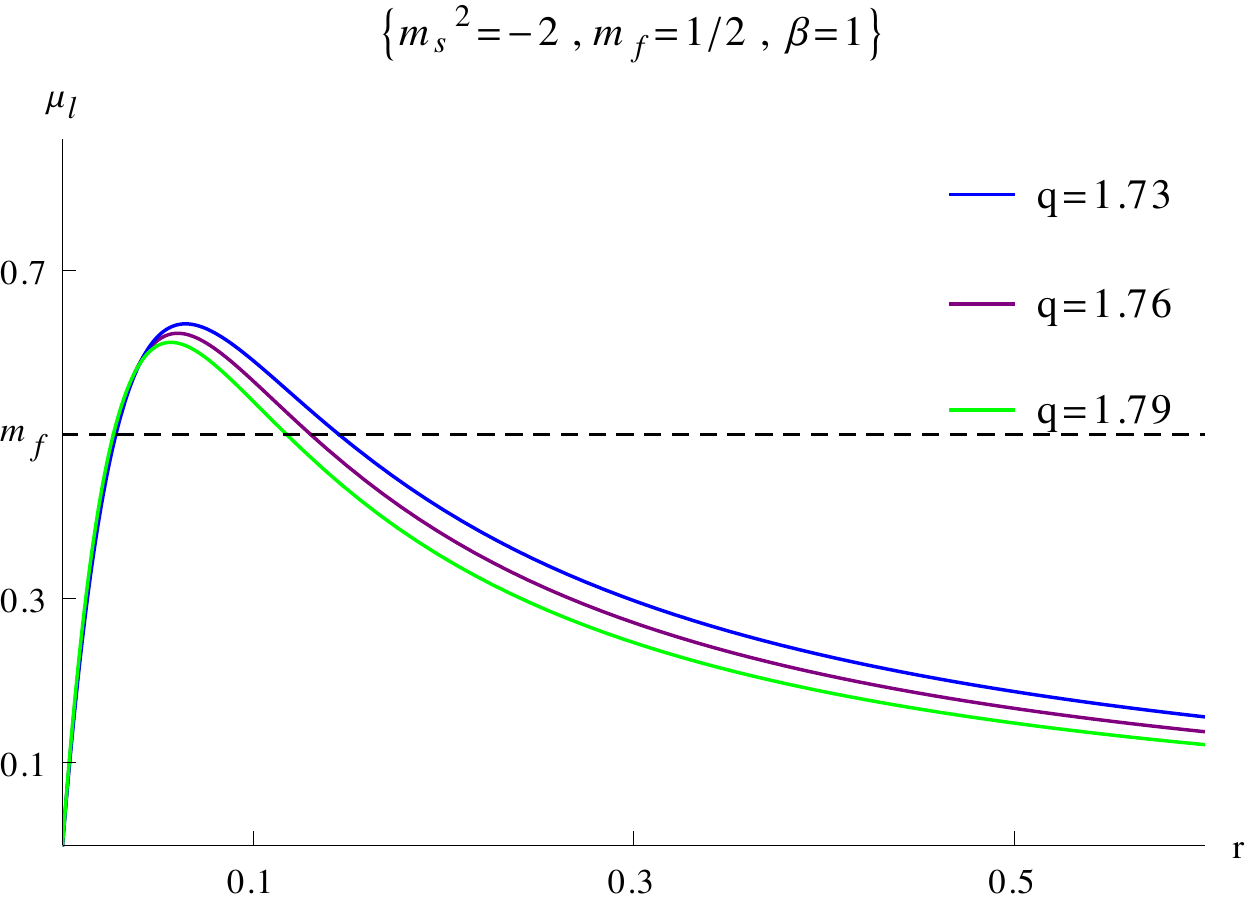}}
\subfloat[ ]{
\includegraphics[width=0.49\textwidth]{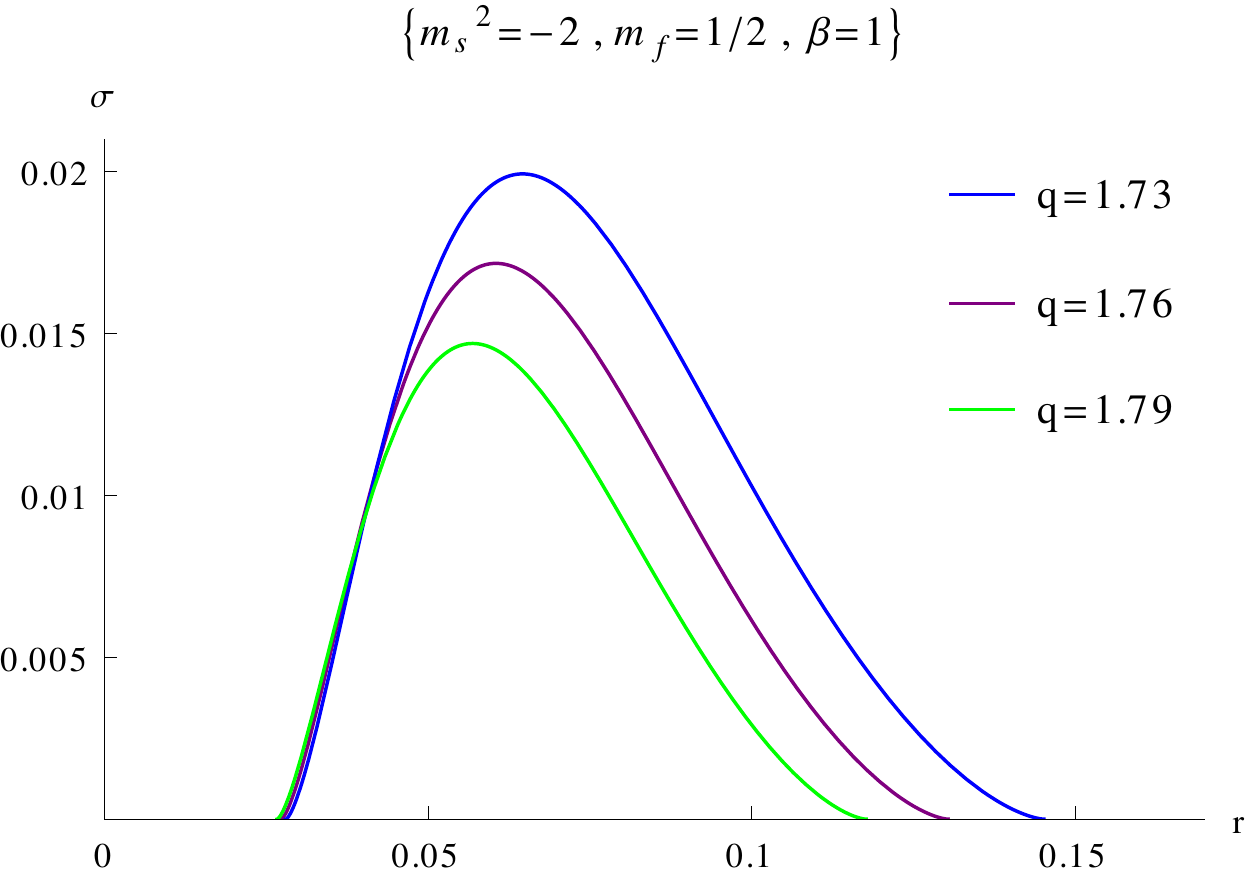}}
\caption{ Profile of (a) the local chemical potential and (b) the charge density for different solutions of the compact electron star, with fixed $m_f=0.5$ and $\beta=1$,  built starting from the superconductor solution. The star is confined to the region where $\mu_l(r) > m_f$.}
\label{CSfig}
\end{figure}

\begin{figure}[h!]
\begin{center}
\includegraphics[width=0.49\textwidth]{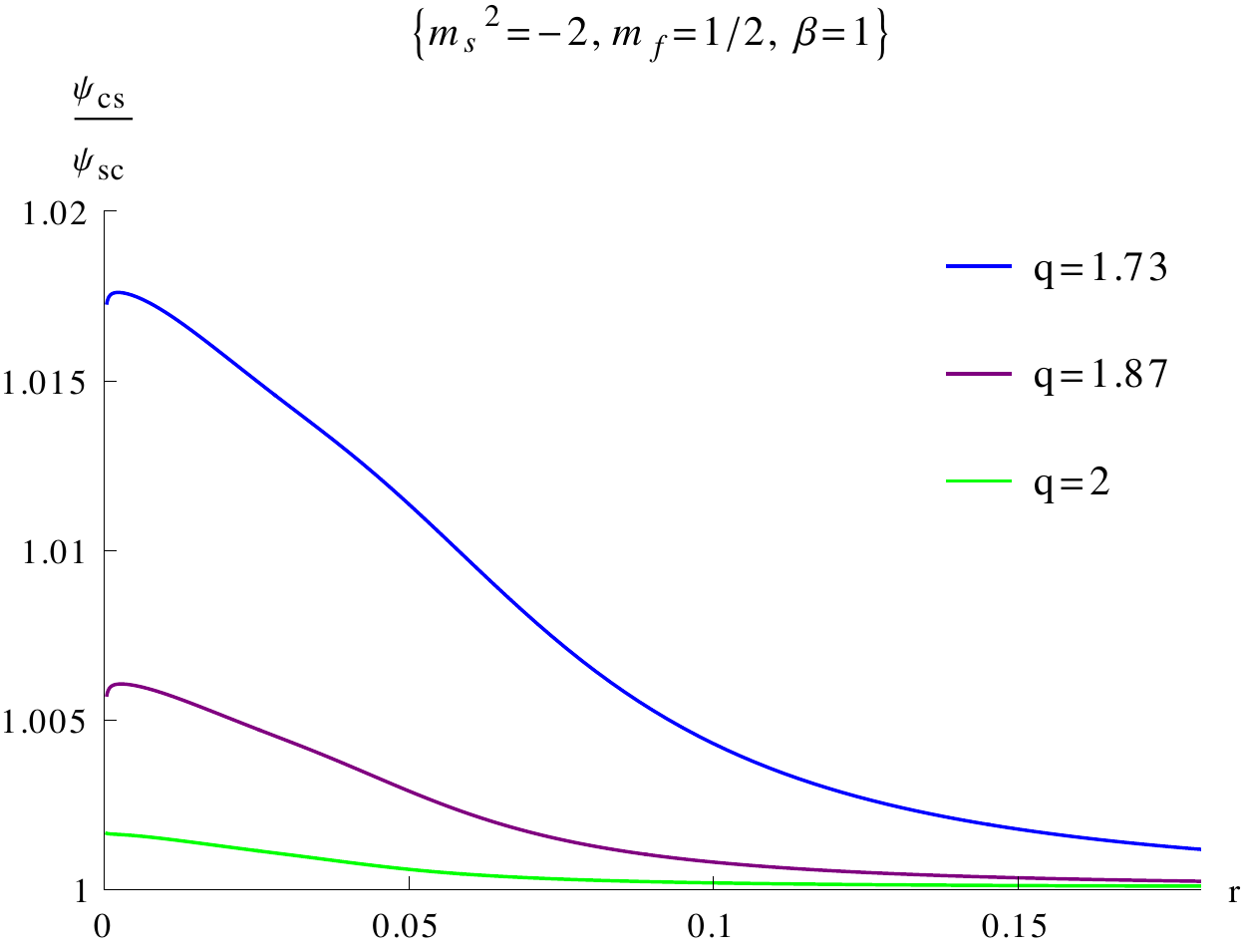}
\caption{ Profile of the condensate of the scalar field in the bulk in the compact star over its value in the holographic superconductor for different values of the scalar field charge $q$. }
\label{CSfig2}
\end{center}
\end{figure}

This situation is displayed in Figures \ref{CSfig} and \ref{CSfig2}, where we show the local chemical potential profile of some typical solutions (found numerically), the compact star density profiles and the values of the scalar field. 

%

In this solution, the charge is shared between the scalar condensate and the fluid, $Q = Q_{s} + Q_{f}$ , with both components non-vanishing. 
The phases ES, SC and CS are all cohesive (in the dual language), since for these solutions the electric flux vanishes at infinity in the IR, so the charge in entirely in the bulk degrees of freedom. 

The compact star exists under the condition 
\be\label{ir1}
m_f < \mu_{max}(q,m_s)\,,
\ee 
where $\mu_{max}(q,m_s)$ is the maximum value for a given SC solution with parameters $(q,m_s)$. This condition is independent of the  total charge $Q$ of the solution as will become clear in the next section.
\begin{figure}[b]
\centering
\subfloat[ ]{\includegraphics[width=0.49\textwidth]{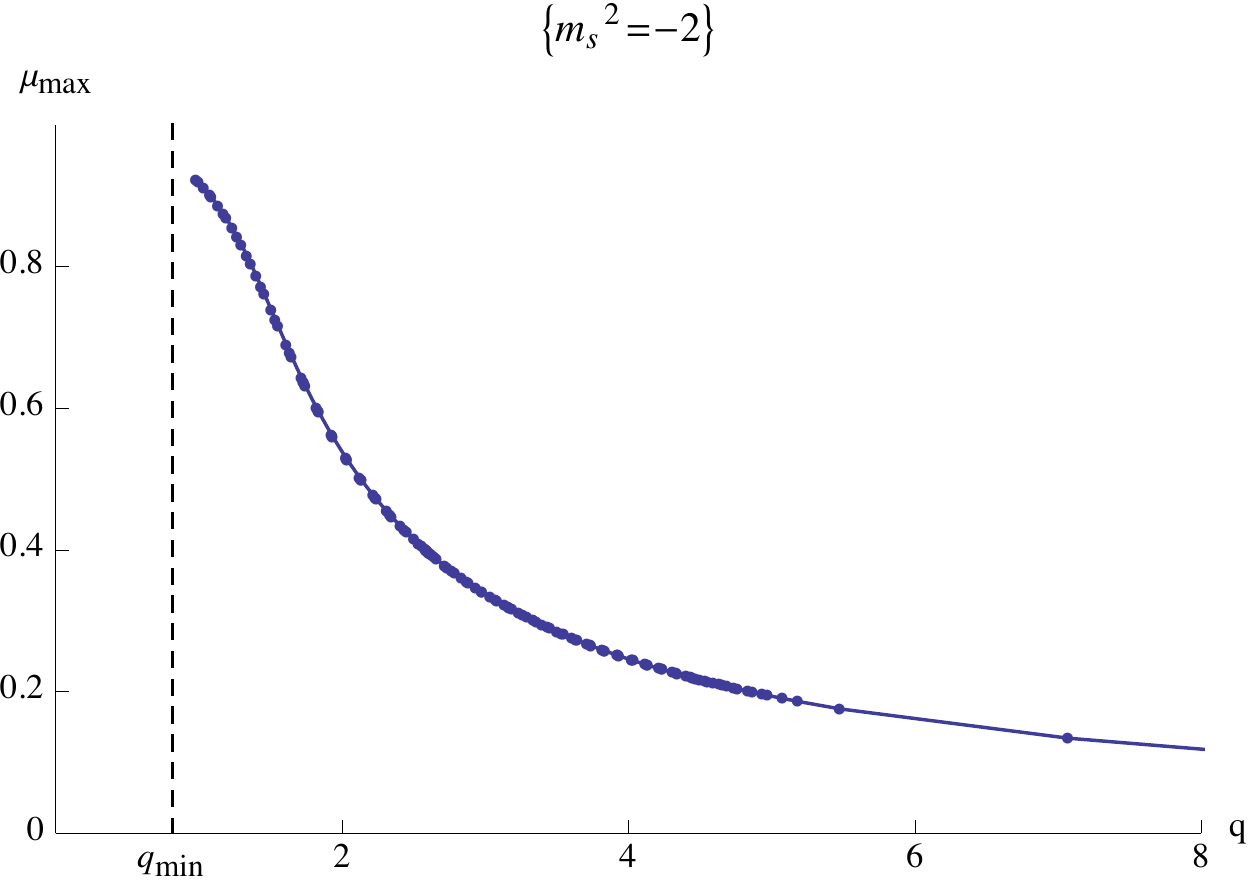}}
\hskip0.1cm
\subfloat[ ]{\includegraphics[width=0.49\textwidth]{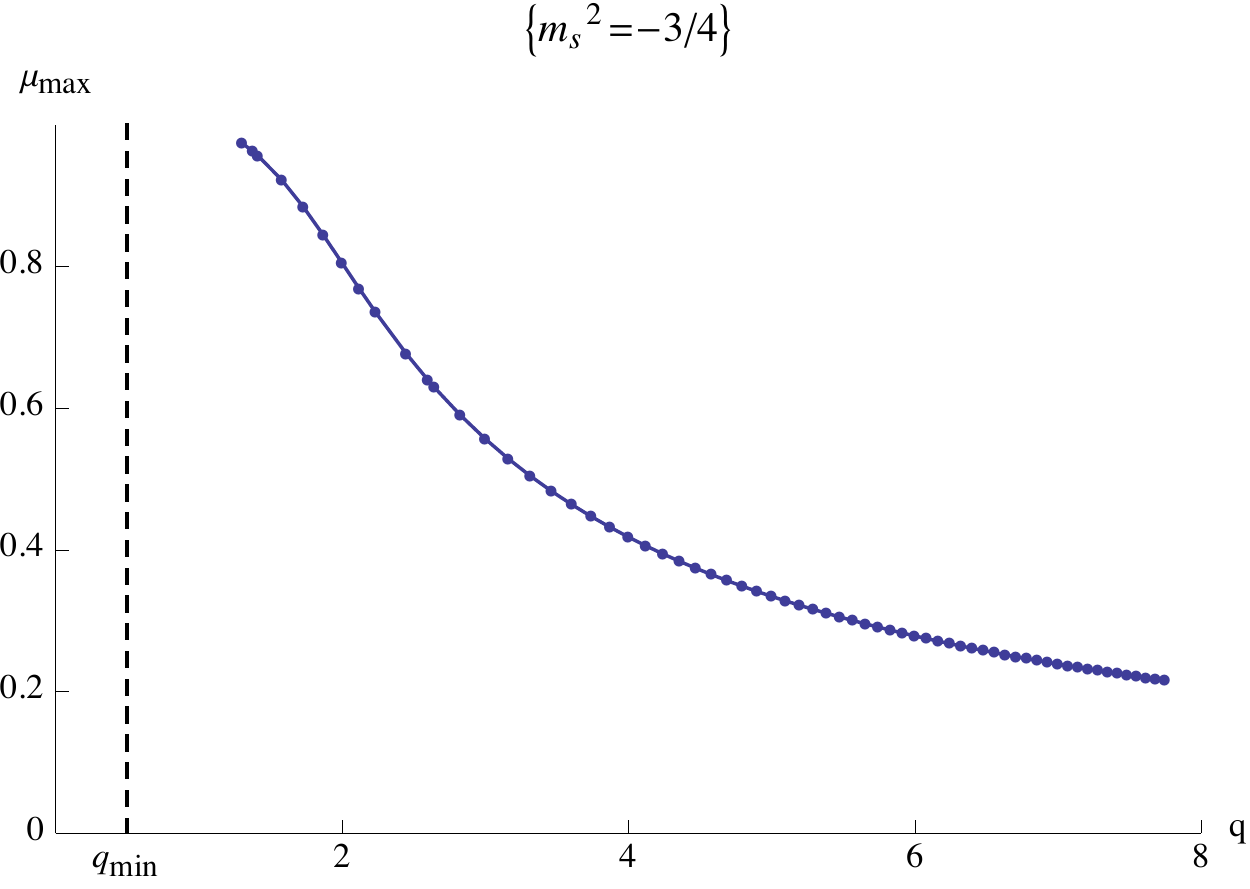}}
\caption{Maximum local chemical potential $\mu_{max}$ reached in  the superconductor solution, as a function of $q$, for (a) $m_s^2=-2$ and (b) $m_s^2=-3/4$.  The compact  star solutions exist in the region below the curve connecting the data points.}
\label{Esistencefig}
\end{figure}
It is very hard to have an analytic handle of the condition (\ref{ir1}) as a function of the parameters. Finding the solutions  numerically and  scanning the parameter space, we have found the region of existence of the compact star, displayed in Figure \ref{Esistencefig}. The curve should extend to the point $q=q_{min}$ with  $q_{min}^2 \equiv-m_s^2/3$. Below this point the IR geometry in (\ref{sc6}) is not correct. However the numerics become very hard to control for values of $q^2$ smaller than unity.  Also, we have found that $\mu_{max}$ is always smaller than one, thus when the compact star exists, it must have $m_f<1$, just as for the unbounded star.

\begin{figure}[t]
\centering
\subfloat[ ]{\includegraphics[width=0.48\textwidth]{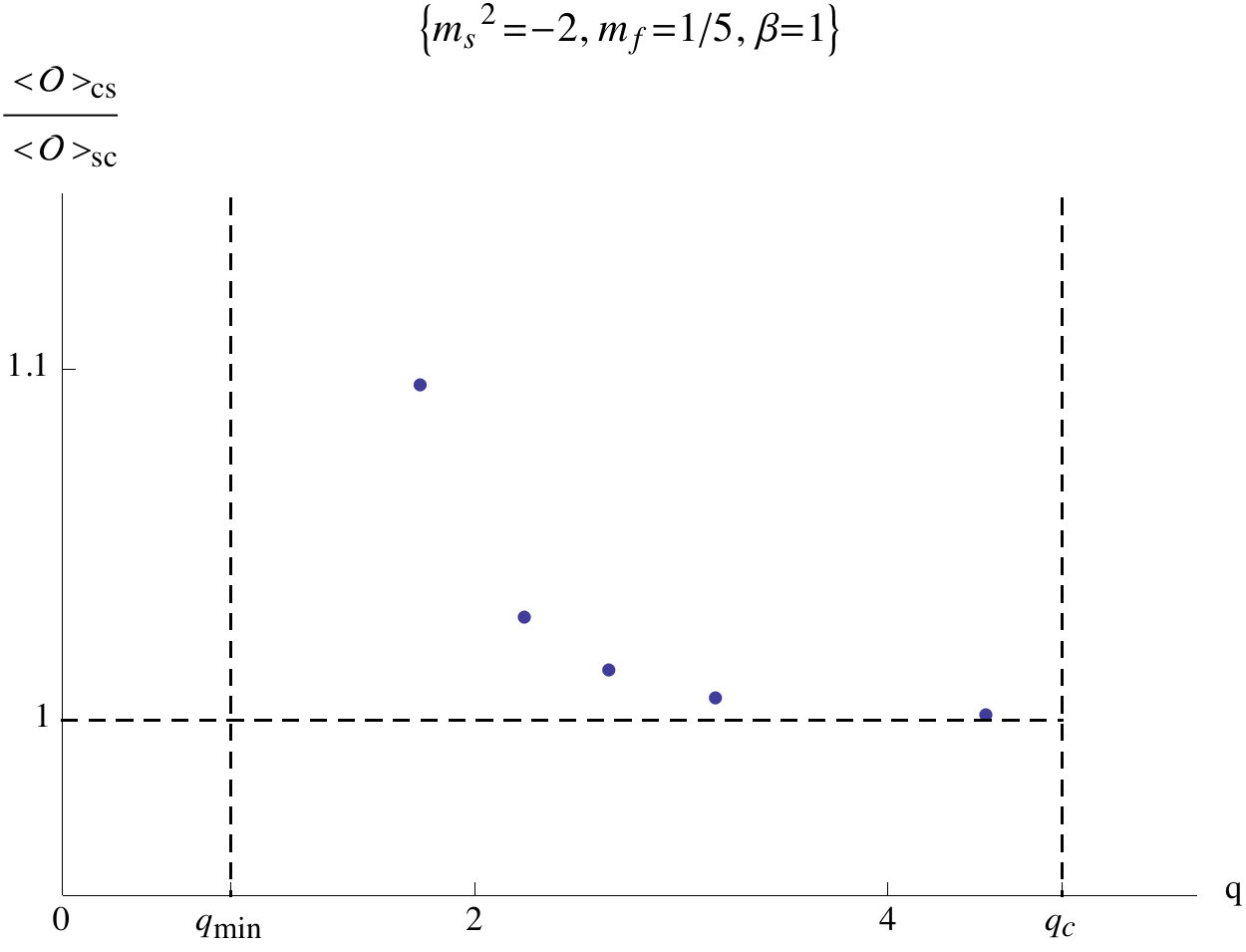}} 
\hskip0.1cm 
\subfloat[ ]{\includegraphics[width=0.48\textwidth]{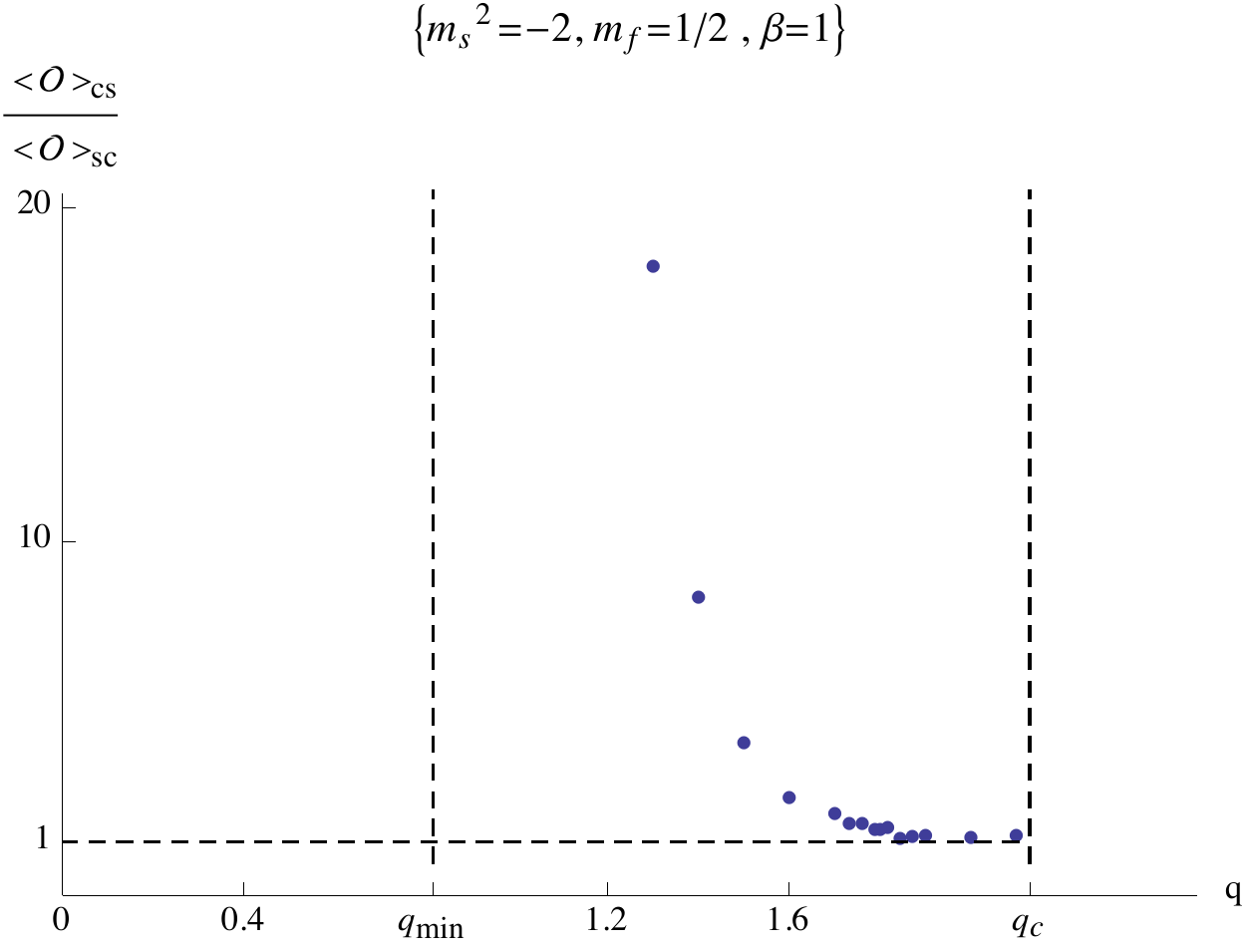}} \\
\subfloat[ ]{\includegraphics[width=0.48\textwidth]{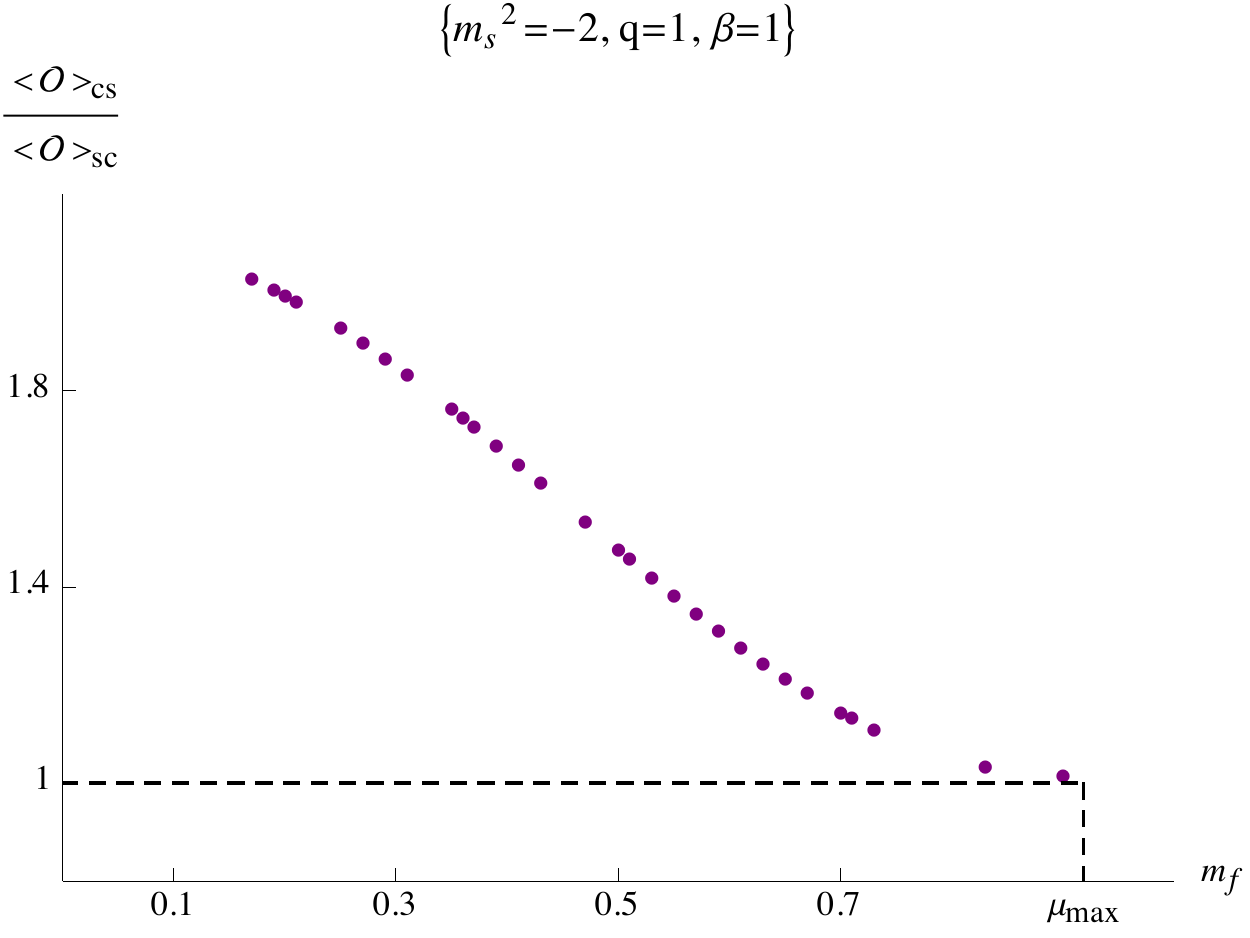}}
\hskip0,1cm
\subfloat[ ]{\includegraphics[width=0.48\textwidth]{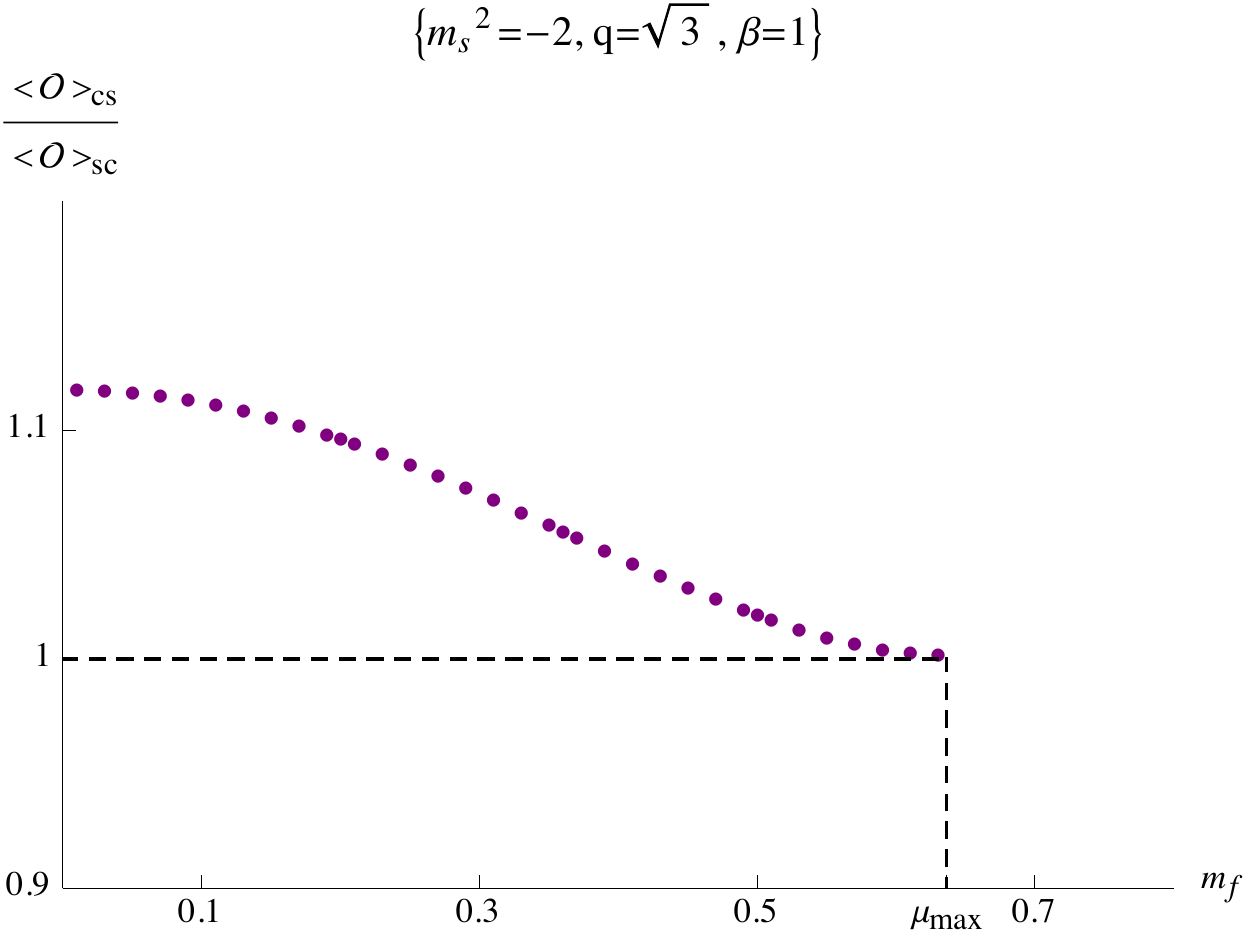}}
\caption{Ratio of the condensate in the compact star solution over its value in the holographic superconductor for the same parameters.}
\label{condfig}
\end{figure}

Figures \ref{condfig}  show the ratio between the scalar condensate $\psi_+ \propto \< O\> $ in the superconductor vs. the compact star solutions  as functions of $q$  and $m_f$. The presence of the star appears to increase the condensate from the SC value.

\subsection{Solution-generating symmetries and physical parameters} \label{scaling}

Having found four different branches, it is important to understand what controls  the solutions on a given branch, once the parameters in the Lagrangian are fixed. To understand this issue, we need to count the possible deformations  of the system (in particular, its IR geometry) and put them in relation with physical parameters distinguishing one solution from the other in the same branch.

In this regard, it is useful to notice the existence of two independent symmetries of the field equations:
\begin{subequations}\label{symm}
\begin{align}
 (r,x,y)\to a(r,x,y) \ , \ \ &\ \ f \to a^{-2}f \ , \ \ \ \ g\to a^{-2}g \ , \ \ \ \ h\to
a^{-1}h \ , \label{sym-a} \\ 
&&\nonumber\\
& f\to b^{-2} f \ , \ \ \ \ h\to b^{-1}h \,. \ \label{sym-b}
\end{align}
\end{subequations}
Notice that these are {\em not} symmetries of the ansatz (\ref{rec1}): the metric and gauge field are not invariant, and we cannot undo the transformation by a time diffeomorphism. Rather, (\ref{symm}) are  solution-generating symmetries, which take one solution to a physically different one  (e.g. with different mass and charge).  

\subsubsection*{Extremal RN branch}
First let us consider the ERN branch: in this case, clearly the only parameter of the solution (if we want to remain at extremality) is $r_+$, or equivalently the chemical potential. Changing $r_+ \to r_+/\lambda $ will scale the mass, chemical potential and charge by 
\be \label{sym1}
(\mu,Q,M) \to (\lambda\mu, \,\lambda^2 Q, \,\lambda^3 M). 
\ee
This operation can be achieved by performing two  consecutive transformations 
of the type (\ref{symm}), with $a=\lambda^{-1}$ and $b=\lambda$. Thus, we can generate by simple scaling all extremal black hole solutions with different charges. Although this is not obvious, it turns out that the same is possible for all other branches as well.

\subsubsection*{Electron Star branch}

In this case, the parameters $m_f$ and $\beta$ fully specify the Lifshitz exponent and the coefficients of the asymptotic geometry (\ref{star4}). As discussed in \cite{Hartnoll:2010gu},  the solution admits a single UV-relevant deformation, parametrized by the coefficient $f_1$: 
\begin{equation}
\label{sym2}
\begin{aligned}
f &= \frac{1}{r^{2z}}\left(1+f_1 r^\alpha + \dots \right) \ ,  \ \ \ \ g =
\frac{g_\infty}{r^2}\left(1+g_1r^\alpha + \dots \right) \ , \\
h &= \frac{h_\infty}{r^z}\left(1+h_1r^\alpha + \dots \right) \ ,  
\end{aligned}
\end{equation}
where $\alpha$ is determined by the parameters, and $g_1$ and $h_1$ are functions of $f_1$. There are no other deformations: they either destroy the UV geometry, or introduce a temperature. Thus,  changing $f_1$ can only generate  different star solutions with different charges. 

However, changing $f_1$ can be also achieved by a symmetry transformation of the kind  (\ref{sym-a}) , followed by one of the kind (\ref{sym-b}) to restore the right normalization of the leading term of $f(r)$ in the UV, i.e. to ensure that $f\sim 1/r^2$ as $r\to 0$. Thus, stars with different charge and chemical potentials are connected by the symmetry transformations, which
imply simple scaling transformations like (\ref{sym1}) on the parameters of the solutions.     
  
\subsubsection*{Superconductor  and Compact Star branch}

The  same considerations apply to the  superconductor and the compact star, which can be discussed together since they have the same IR asymptotics. 
In eq. (\ref{sc6}) the parameter $h_0$ is not free, because it has to be fixed in such a way that the source term in the UV expansion of the scalar field vanishes. We can identify a deformation parameter by noticing that replacing
\be\label{sym4}
\log r \to \log r/r_0
\ee
for an arbitrary $r_0$, eq. (\ref{sc6}) are still an asymptotic solution for large $r$. Thus, we can change $r_0$ to obtain different solutions in this branch. On the other hand, we can remove $r_0$ by a combination of the symmetries (\ref{symm}) which involves a rescaling $r\to r/\lambda$. As it is clear from the UV asymptotics (\ref{uvasym}),   the new solution will be related to the old one, up to subleading corrections,  by a simple scaling of the mass, charge, chemical potential {\em and} value of the condensate: 
\be\label{sym5}
(\mu,Q,M,\psi_+) \to (\lambda\mu, \,\lambda^2 Q, \,\lambda^3 M, \, \lambda^{\Delta}\psi_+) \,.
\ee 
The conclusion  of this subsection is that, on each branch, the transformations
(\ref{symm}) connect solutions with different values of the  charge and chemical potential, and these will obey simple scaling relations as (\ref{sym1}).
 
\section{Free energy and the phase diagrams}\label{phase}

In the holographic dictionary, the free energy of a state corresponding to a gravitational solution is computed by the on-shell action of that solution. At zero temperature, the only parameter is the chemical potential, so we want to obtain the free energy as a function of $\mu$  in the grand canonical ensemble. 

For all solutions, the on-shell action  per unit boundary 2d volume $S$, mass density $M$ and charge density $Q$  of the solution correspond to the free energy per unit volume $F$, energy density and charge density on the boundary side respectively, and  as we have seen  in Section \ref{scaling}, the solutions on each branch depend on a single parameter. In the grand canonical ensemble the independent parameter is taken to be the chemical potential $\mu$.  The thermodynamic quantities  are related by the first law
\be\label{first1}
F(\mu) = E(\mu)-\mu Q(\mu).  
\ee 
We can check the validity of eq. (\ref{first1}) by computing separately the on-shell action, ADM mass  and total  electric flux   of the solution in each of the four branches we are considering. The calculation is described in Appendix \ref{AppOnshell}. In fact, once these quantities are  written as  UV boundary terms, the calculation reduces  to the case of the extremal black hole, where the solution is exactly given by (\ref{bh1}) : the reason is that both the star  and the condensate  deform the solution in the UV only by 
terms which are subleading, and do not contribute at all to any of the quantities in eq. (\ref{first1}): for the star, this is obvious since the exterior solution coincides with the extremal black hole; for the condensate and the compact star, this is a consequence of the asymptotic behavior (\ref{uvasym}) and in particular the fact that $f(r)$ and $h(r)$ are unchanged to the order that gives finite contributions to eq. (\ref{first1}). 

Next, we must determine the equation of state of the solutions, i.e. the functional  relation between $Q$ and $\mu$.  Due to the simple scaling relating the solution on any branch, see eq.  (\ref{sym1}) and (\ref{sym5}), the relation between the charge and the chemical potential is always quadratic, with the only unknown being the coefficient:
\be\label{free1}
Q_i(\mu) = c_i \mu^2, \qquad i=1\ldots4 \,, 
\ee 
where $i$ labels the various branches.

In a grand-canonical ensemble, at zero temperature, we have the thermodynamic relation
\be\label{free1a}
Q = -{d F\over d\mu} \,.
\ee
By integration, and by fixing the zero charge solution to have zero free energy (it is in all cases the pure $AdS_4$ solution), we obtain the simple conformal result for the free energy on each branch:
\be\label{free1b}
F_i (\mu) = -{1\over 3} c_i \mu^3 \,. 
\ee 
Inserting again eq. (\ref{free1b}) into the first law leads  to $E = -2F$, i.e. {\em a conformal equation of state}, no matter on which branch we are. This can be checked independently by computing the holographic stress tensor and verifying that it is traceless. This is also done in Appendix \ref{AppOnshell}, where we find that indeed the renormalized holographic stress tensor is traceless, which justify the use of  (\ref{free1a}) to compute the free energy as a function of $\mu$. 
Again, the computation reduces to the computation in the extremal  black hole, as the deformations caused by the star and the condensate do not change any of the  leading terms in $f(r)$ and $g(r)$ giving a finite contribution at the UV boundary $r\to 0$. 

The coefficients $c_i$ are features of each branch, and depend only on the parameters of the model, and not on the charge of the solution. Thus, the phase diagram as a function of $\mu$, or $Q$, is trivial: a solution which is favored for one value of $\mu$, will be favored for all values. Which branch has the largest free energy is decided only by value of the coefficients $c_i$. 

For the extremal black hole, the coefficient can be extracted from eq. (\ref{bh2}), 
\be
c_{ERN} = {1 \over \sqrt{6}} \, , 
\ee 
but for the other branches it must be computed numerically.  

In order to compare the different branches, the solutions have to be found numerically. We used a numerical procedure based on Mathematica NDSolve, shooting from the IR at a large cut-off, where we impose the right IR asymptotics,  and then matching the solution to its correct UV behavior (normalization, absence of sources). In the solution involving the fluid, 
we impose continuity of the functions and their derivative at the star boundaries. 
 
We checked the relation (\ref{free1}) numerically on the various branches, and we found it is obeyed with great accuracy on a large range of values of $\mu$ and $Q$. This both confirms the validity of (\ref{free1}),   and constitutes a check  of our numerical procedure. 

We can then describe our results for the phase diagram of the system. To reiterate, there are 4 solutions:
\begin{enumerate}\item 
 the black hole, that exists for all values of the parameters; 
\item  the electron star that exists for $m_f < 1$; 
\item  the superconductor, that exists in the region given by (\ref{sc2}); 
\item   the compact star, that exists in the region (\ref{ir1}).
\end{enumerate}
In addition to these constraints, we must stress that for $q^2<|m_s^2|/3$ but $m_s^2-q^2<-3/2$, although  a  non-trivial solution for  the superconductor (and thus also for the compact star) should exist, its form is not known at present. Thus, this region will be outside the reach of our investigation. 

\begin{figure}
\centering
\subfloat[ ]{\includegraphics[width=0.49\textwidth]{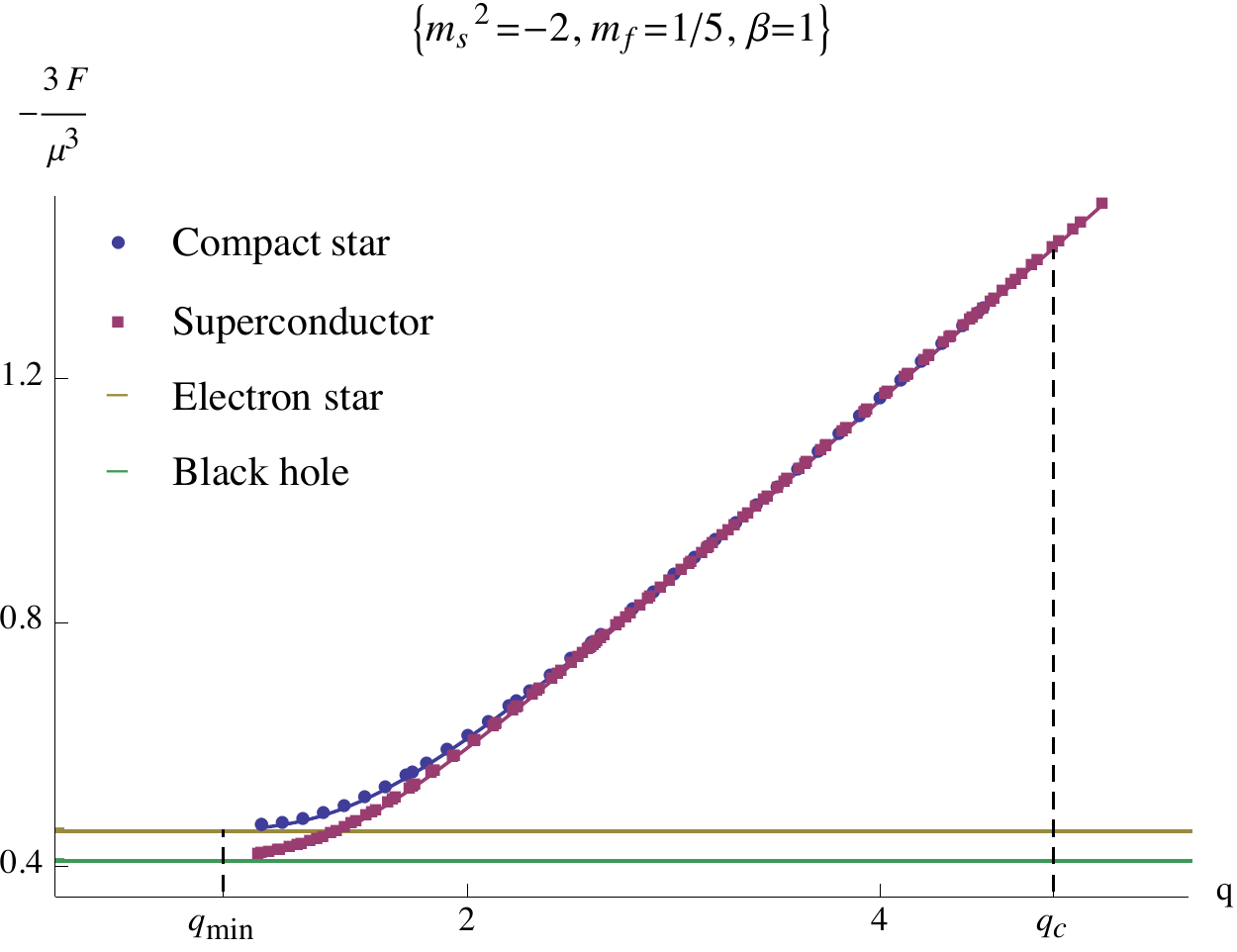}}
\subfloat[ ]{\includegraphics[width=0.49\textwidth]{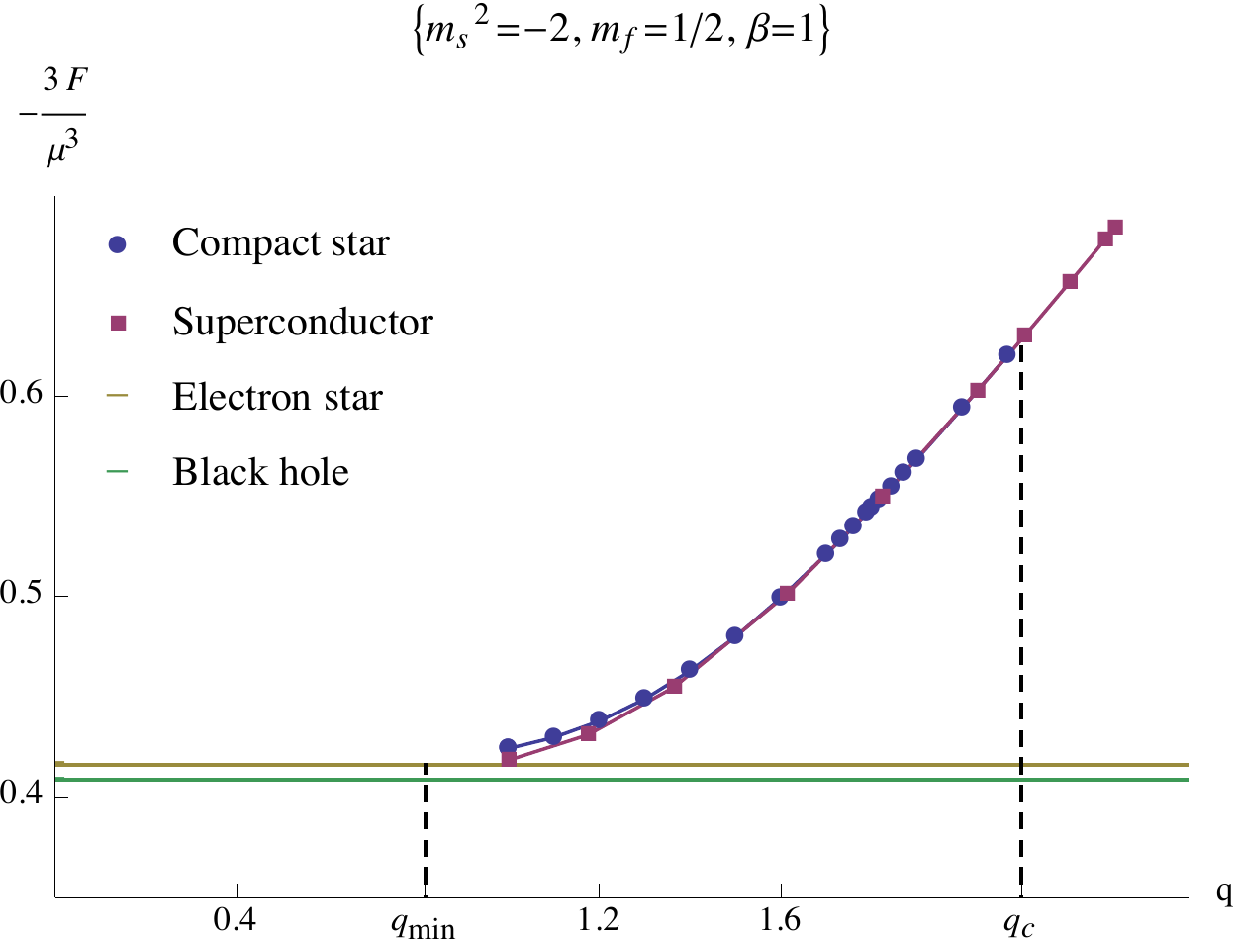}}\\
\subfloat[ ]{\includegraphics[width=0.49\textwidth]{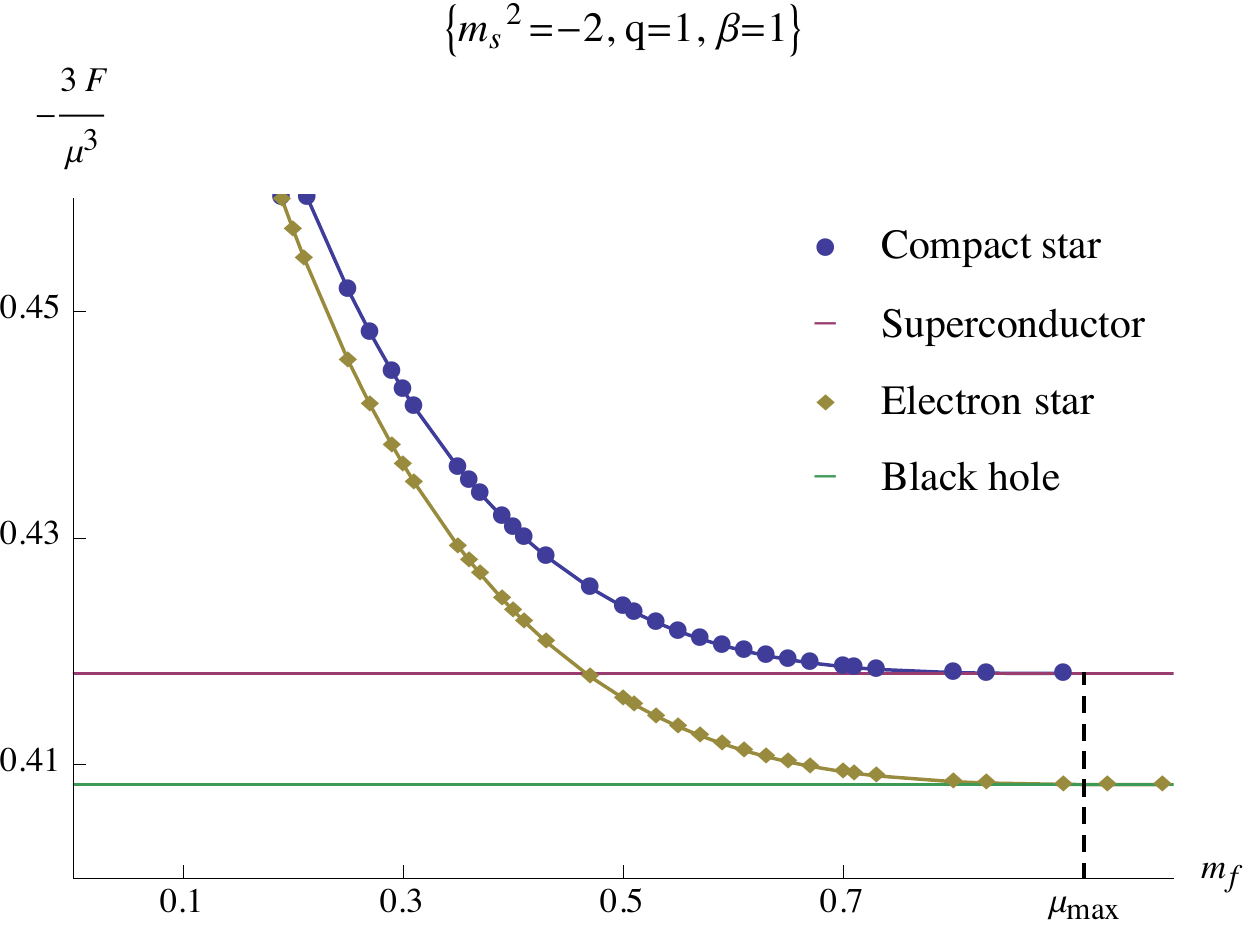}}
\subfloat[ ]{\includegraphics[width=0.49\textwidth]{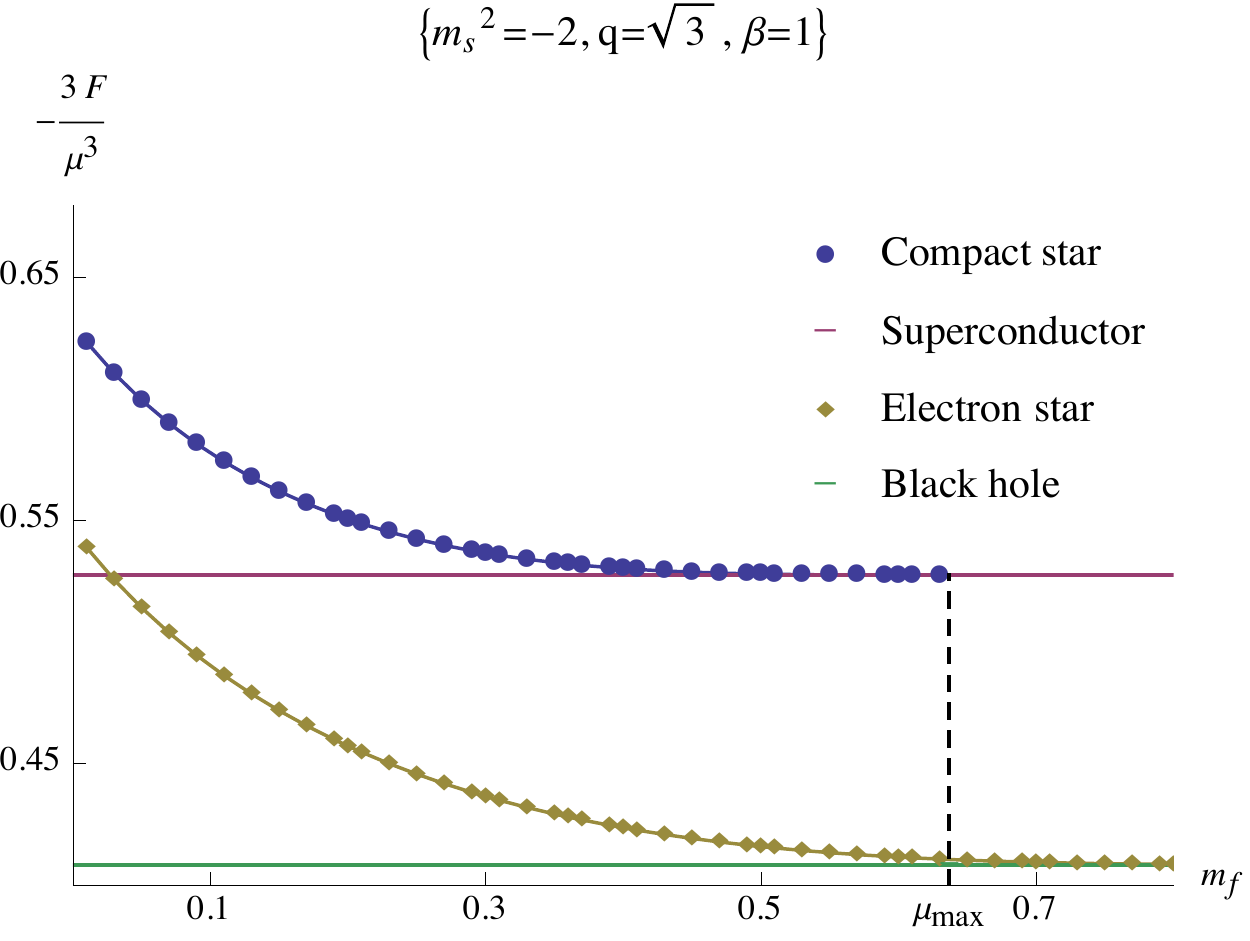}}
\caption{Free energy (normalized to the chemical potential)  of the four competing solution, for fixed scalar mass $m_s^2=-2$. The lines interpolating the data points are drawn for visual aid. In (a) and (b) $F$ is plotted as a function of $q$ for two fixed values of the fermionic mass $m_f$. The superconductor and the compact star solution are unknown for $q<q_{min}$, whereas the electron star and the black hole continue past this point. In (c) and (d) it is a function of $m_f$ for two fixed values of the scalar field charge . The  compact star solution exists only for $m_f< \mu_{max}$, where it merges with the superconductor solution. The electron star merges with the black hole solution as $m_f\to 1$. }
\label{fig:free}
\end{figure}

\begin{figure}
\centering
\subfloat[ ]{\includegraphics[width=0.49\textwidth]{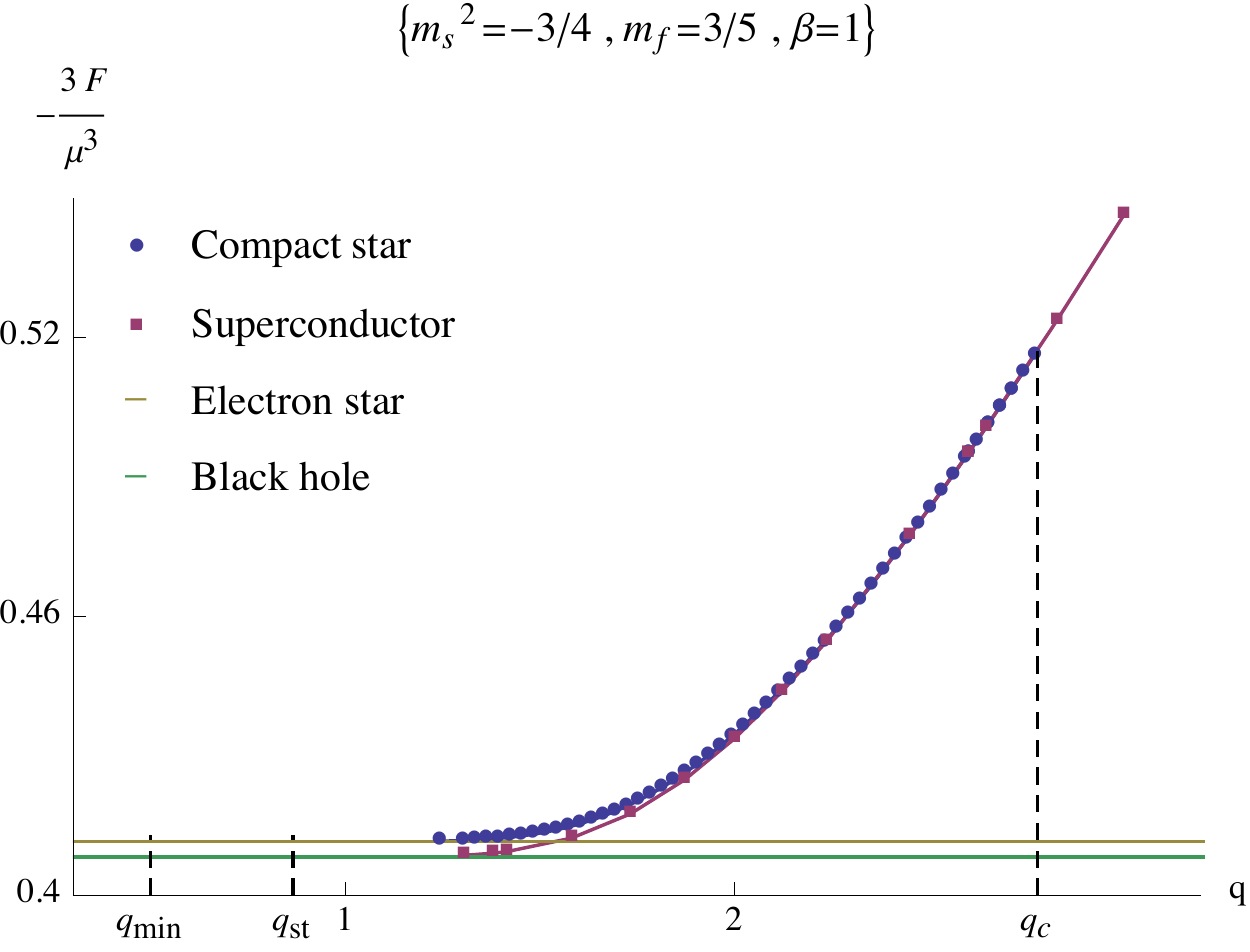}}
\subfloat[ ]{\includegraphics[width=0.49\textwidth]{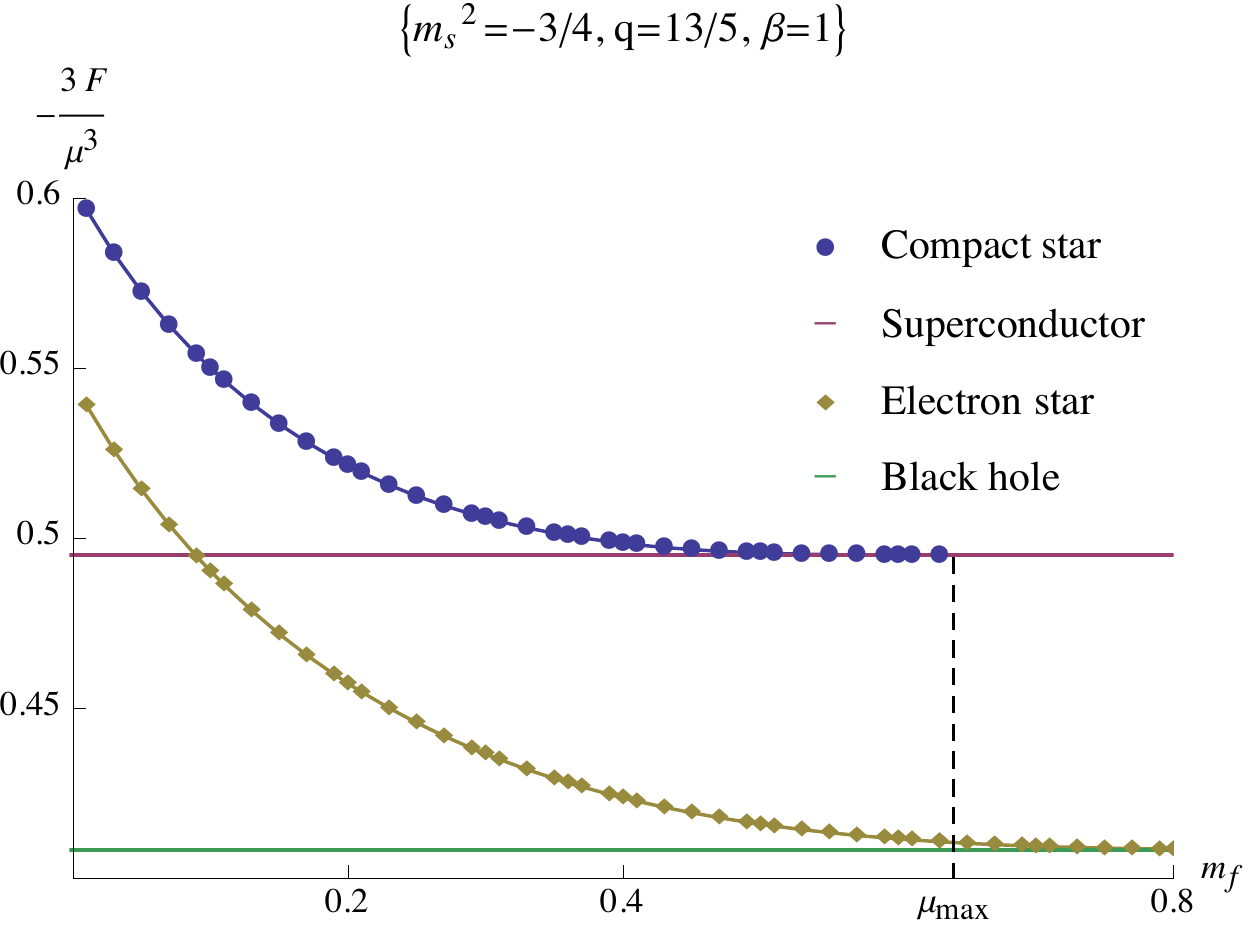}}
\caption{Free energy (normalized to the chemical potential)  of the four competing solution, for fixed scalar mass $m_s^2=-3/4$, (a) as a function of $q$ for a fixed value of $m_f$, (b) as a function of $m_f$ for a fixed value of $q$. The lines are again merely for visual aid.}
\label{fig:free-other}
\end{figure}

We computed numerically the free energy as a function of $q$ and $m_f$, for $m_s^2 = -2$ and $m_s^2=-3/4$. The resulting curves are displayed in Figure \ref{fig:free} and \ref{fig:free-other}. 

We found that the compact star solution is the favored solution in the region where it exists. There is a crossing of the SC and ES branches but, as far as we could determine, it is always in the region where the CS solution is favored so it does not correspond to a phase transition. 

\begin{figure}
\centering
\subfloat[ ]{\includegraphics[width=0.49\textwidth]{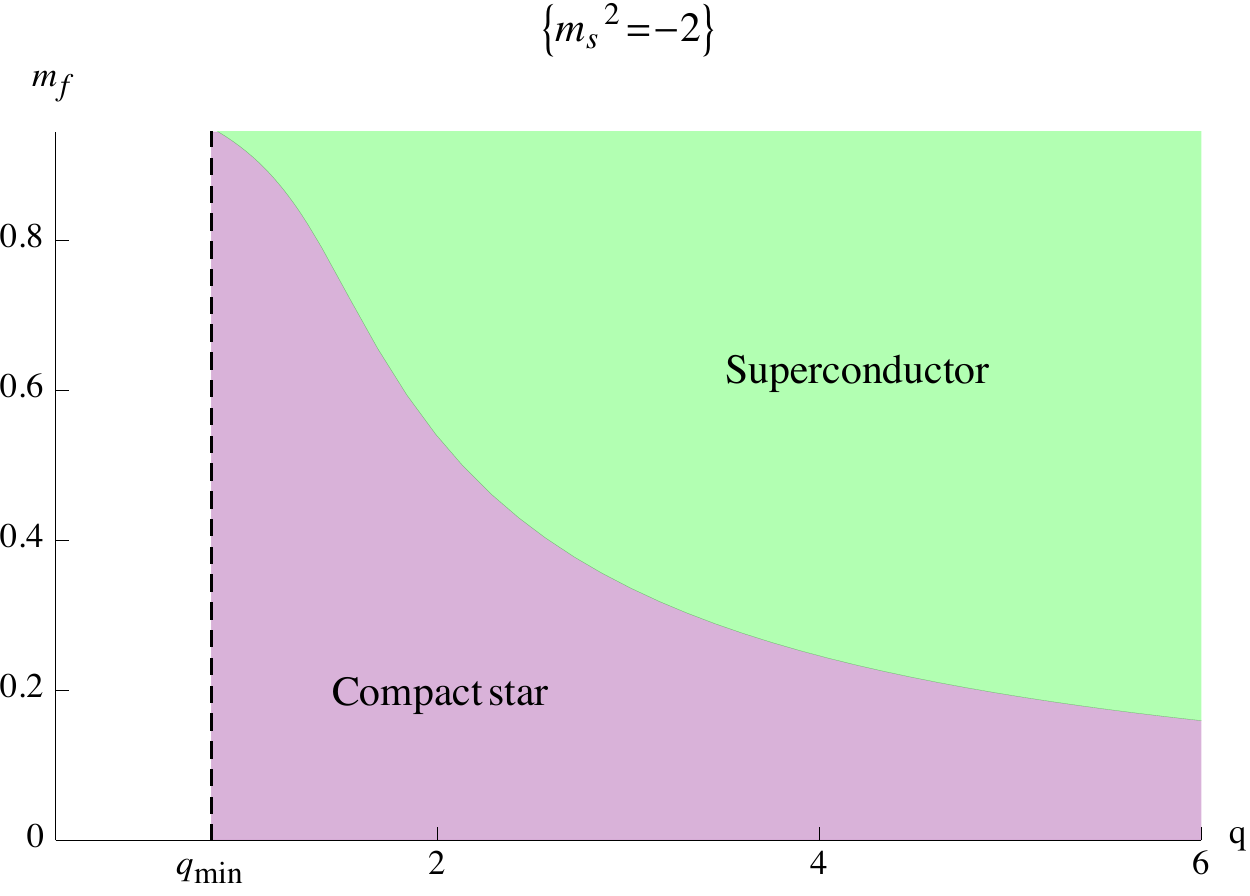}}
\hskip0.1cm
\subfloat[ ]{\includegraphics[width=0.49\textwidth]{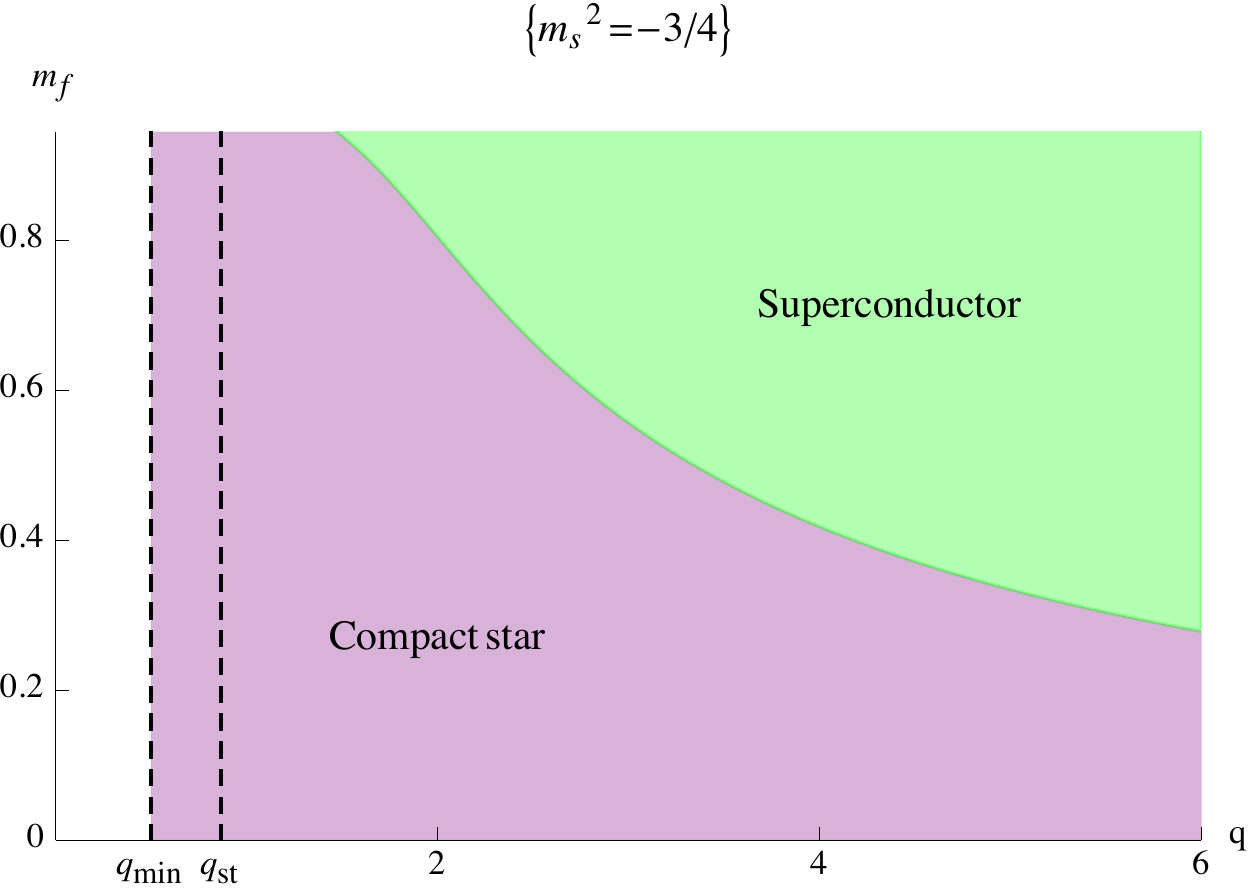}}
\caption{Phase diagram showing the transition between the SC and the CS solutions.}
\label{fig:phase}
\end{figure}

There is instead a phase transition at the point where the CS branch ceases to exist and it connects to the SC solution (at $q=q_c$ for fixed $m_f$). 
The transition appears to be of continuous type as a function of both $m_f$ and $q$, as can be seen from the figures. This is natural to expect if, as it seems to be the case, the compact star starts dominating at the point where it is allowed as a solution, i.e. on the curve $\mu_{max}=m_f$, where it has the same free energy as the pure superconductor solution. As a result we obtain the phase diagram of the system shown in Figure \ref{fig:phase} again for the cases $m_s^2=-2$ and $m_s^2=-3/4$. 

As we have mentioned earlier, the branches with non-zero condensate are known only for $q^2 > - m_s^2 /3$, and this is the range in which our ansatz gives a solution. For smaller values of $q$ the ES and ERN solutions are unstable  
so there should be other solutions with non-zero condensate but they are not known at present, so we cannot draw conclusions about the phase diagram but we can make some conjectures. In \cite{Horowitz:2009ij} two putative additional branches of solutions with condensate were found in the region $0 < q^2 < - 2 m_s^2/5$ but they were not able to determine whether these solutions have free tunable parameters to allow them to connect to the UV asymptotics. At any rate one of the two branches seemed to connect smoothly onto the 
region $q^2> - m_s^2/3$. Another source of uncertainty is that our level of numerical precision does not allow us to determine whether the SC matches the BH solution precisely at the boundary of this region. So we are left with the following possibilities: 

1) the SC branch  connects smoothly  at $q_{min}$ onto another branch that extends to $q \to 0$ and dominates over the ERN solution everywhere; 
2) the SC branch stops at some point (perhaps not equal to $q_{min}$), and there are two other branches 
going down to $q=0$; this implies that there should be another phase transition, probably of first order, between two different types of superconductors. 
The same should be true regarding the fate of the compact star vs. the electron star. It would be very interesting to find solutions in the region $q^2 < - m_s^2/3$ in order to settle the issue. 

In Figures \ref{fig:free-other} and \ref{fig:phase} we have also marked the point $q=q_{st}$ where the scalar mode around the ERN becomes unstable, determined by eq. (\ref{sc2}); for 
 $m_s^2 > -9/8$ (and so in particular for the case we analyze $m_s^2=-3/4$) one finds that $q_{st} > q_{min}$. The plot indicates that the CS solution is crossing the ES around that point but we cannot determine 
if there is really a crossing point before reaching $q_{min}$. If there is one between
$q_{min}$ and $q_{st}$ then there is another phase transition between CS and ES; if moreover
the crossing happens for $q>q_{st}$ then there would be a region where neither the ES nor the
CS is the dominant solution. We hope to return to this question in the future with improved
accuracy.

\section{Discussion}\label{discussion}

We have found new solutions, that we called Compact Stars, in which the charged matter sourcing the geometry includes a fermionic fluid and a scalar with non-vanishing condensate. 
We have characterized this solution with respect to the other known ones with the same asymptotics and found the phase diagram. We could determine the boundary between the CS and the SC phase. 
Within the limits of our numerical accuracy we could not definitely rule out the presence of other boundaries though we suspect the presence of another one between the CS and the ES. 

We conclude by pointing out some open questions that would be interesting to investigate in the 
future. 

Regarding the solutions that we have found, we would like to be able to determine the order 
of the phase transition; perhaps this could be done analytically with arguments along the lines of those presented in \cite{Hartnoll:2010ik}. 
We should consider fluctuations of the gauge field, in order to determine the conductivity, and of probe fermions, in order to detect the Fermi surface and explicitly show how it is affected by the fractionalization. 
In general we would like to have a better understanding of the significance of the transition from the boundary theory point of view. 

A natural extension would be to consider the phase diagram with more parameters (e.g. temperature, 
magnetic field); one could also add angular momentum if the space is asymptotically global AdS \cite{Sonner:2009fk}. 
Probably the most interesting direction is to try and find solutions with non-vanishing condensate at  small $q^2$ in order to ascertain the true phase of the system in that region. 

Physically we expect also that the fluid, being ultimately composed of charged elementary fermions, could also have direct coupling to the boson, and not only via exchange of photons. If one does not want to leave the fluid approximation, the simplest  possibility would be a current-current interaction.

\subsection*{Acknowledgements}
We would like to thank Blaise Gout\'eraux and especially Koenraad Schalm for useful exchanges and discussion. 
  
\addcontentsline{toc}{section}{Appendices} \appendix

\renewcommand{\theequation}{\Alph{section}.\arabic{equation}}

\section{Action,  field equations, conventions} \label{Fieldeq}
In this appendix we give some of the details about the action, field equations and parameters of the model under consideration, and about how our conventions are related to those found in the literature.  We take the  signature of the metric to be  $(-,+,+,+)$. Our conventions match those of the electron star  paper \cite{Hartnoll:2010gu}, and differ by those of the zero-temperature superconductor paper \cite{Horowitz:2009ij} by the relative normalization of the gauge field with respect to the Einstein and Scalar sectors. To switch to the notation of ~\cite{Horowitz:2009ij} one has to 
redefine: $A_a\rightarrow(1/\sqrt{2})A_a$ and $q\rightarrow \sqrt{2}\, q$.

The model consists of two sectors:
\begin{enumerate}
\item The Einstein-Maxwell system coupled to a charged scalar, described by the action
\begin{equation}
S = \int \D^4x \sqrt{-g}\left(\mathcal{L}_{\mathrm{Eins.}}+\mathcal{L}_{\mathrm{Mxwl.}}
+ \mathcal{L}_{\mathrm{scalar}}\right)
\end{equation}
where
\begin{equation}
\mathcal{L}_{\mathrm{Eins.}}=\frac{1}{2\kappa^2}\left(R+\frac{6}{L^2}\right) \ , \ \ \
\mathcal{L}_\mathrm{Mxwl.}=-\frac{1}{4e^2}F_{ab}F^{ab} \ ,
\end{equation}
and
\begin{equation}
\mathcal{L}_\mathrm{scalar}=-\frac{1}{2} \left( \left| \nabla\psi-iqA\psi
\right|^2 + m_s^2 |\psi|^2 \right) \ .
\end{equation}
The stress tensors of the scalar and gauge field  are defined  by
\begin{equation}
 T_{ab}^\mathrm{matter} \equiv - \frac{2}{\sqrt{-g}} \frac{\delta
S_\mathrm{matter}}{\delta g^{ab}} \,. 
\end{equation}

Specifically, we have
\begin{align}
 T_{ab}^\mathrm{Mxwl.} &= \frac{1}{e^2}\left(F_{ac}F_b^{\ c}-\frac{1}{4} g_{ab}
F_{cd}F^{cd}\right) \,, \\
\nonumber \\
T_{ab}^\mathrm{scalar} &= \frac{1}{2}\left( g_a^c g_b^d + g_b^c g_a^d -g_{ab}g^{cd}
\right) \left( \nabla_c\psi-iqA_c\psi\right)\left(
\nabla_d\psi^*+iqA_d\psi^*\right) \nonumber  \\ 
& -
\frac{1}{2} m_s^2g_{ab}\psi\psi^* \ .
\end{align}

The  electromagnetic current of the scalar  field is  given by
\be
 J^b_\mathrm{scalar} = -i \frac{q}{2}g^{ab}\left[
\psi^*\left(\nabla_a-iqA_a\right)\psi - \psi\left(\nabla_a+iqA_a\right)\psi^*
\right] \,.
\ee

\item The  fermionic fluid, described by the local equilibrium, zero-temperature equation of state (in the grand-canonical ensemble)
\be
p(\mu) = -\rho(\mu) + \mu \sigma(\mu), 
\ee
with the energy density $\rho$ and charge density $\sigma$ given by 
\begin{equation}
\rho(\mu) = \beta \int_{m_f}^\mu \D \epsilon \, \epsilon^2 \sqrt{\epsilon^2-m_f^2} \ , \ \ \ \sigma(\mu)
= q_f \beta \int_{m_f}^\mu \D \epsilon \, \epsilon \sqrt{\epsilon^2-m_f^2} \ . \label{fluid}
\end{equation}
where $m_f$ and $q_f$ are  the elementary fermion mass and charge, and $\beta$ is a phenomenological parameter which depends on the microscopic details of the fluid. 
In the geometry, the fluid will feel the local chemical potential $\mu\to\mu_l(r)$.
 
The fluid is coupled to the metric and Maxwell field via its stress tensor and current density, defined  by
\be
T_{ab}^\mathrm{fluid} = (\rho+p)u_a u_b + p g_{ab}, \quad  J^a_\mathrm{fluid} = \sigma u^a \ \,,
\ee
where $u^b$ is the fluid velocity field, constraint to obey
\be\label{norm-u}
u^a u_a = -1.
\ee
The local chemical potential is related to the gauge field by
\be
\mu_l = q_f\, u^a A_a \,.
\ee
This is a choice for the model, since we could in principle  allow for a non-zero ``intrinsic'' chemical potential. Concerning this point, see the discussion in \cite{Hartnoll:2010gu}. 
\end{enumerate}
Einstein's equations, Maxwell's equations, and the scalar field equation are, respectively
\bea
&& R_{ab}-\frac{1}{2}g_{ab}R-\frac{3}{L^2}g_{ab} =
\kappa^2\left(T_{ab}^\mathrm{Mxwl.}+T_{ab}^\mathrm{fluid}+T_{ab}^\mathrm{scalar}\right) \ , \label{einstein}\\
&& \nonumber \\
&& -\left(\nabla_a -iqA_a)(\nabla^a -iqA^a\right)\psi+m_s^2\psi = 0 \ \label{eq:eom-psi} , \\
&& \nonumber \\
&& \nabla_a F^{ba}=e^2 \left(J^b_\mathrm{fluid} + J^b_\mathrm{scalar} \right) \,. \label{maxwell}
\eea
We will restrict to the static and spatially isotropic ansatz
\begin{equation}
 \label{ans1}
\begin{aligned}
\D s^2 &= L^2 \left[-f(r)\D t^2 + g(r)\D r^2 + \frac{1}{r^2} \left(\D x^2 + \D
y^2\right)\right] \ ,\\
A &= {e L \over \kappa} h(r) \D t \ , \qquad \psi = \psi(r),  \qquad u^a = (u^t,0,0,0).
\end{aligned}
\end{equation}
The  non-zero component of the fluid velocity is, by eq. (\ref{norm-u}),   $u^t=1/(L\sqrt{f})$. The  local chemical potential is given by
\begin{equation}
 \mu_l =  q_f u^a A_a = q_f {e \over \kappa}\frac{h}{\sqrt{f}} \,. 
\end{equation}
Also, in order to simplify the notation and to reduce the number of parameters, following \cite{Hartnoll:2010gu} we perform the following parameter and field redefinitions:
\be
\hat{m}_s = m_s L , \quad  \hat{\mu_l} = {\kappa \over e} \mu_l , \quad \hat{\beta} = {e^4 L^2 \over \kappa^2} \beta,  \quad \hat{m}_f = {\kappa \over e } m_f. 
\ee
The local chemical potential and equation of state become simply
\be
\hat{\mu}_l = {h \over \sqrt{f}}, \qquad \hat{p} = -\hat{\rho} + {h\over\sqrt{f}} \hat{\sigma} \,, 
\ee
where $\hat{p},\hat{\rho},\hat{\sigma}$ are defined as in eq. (\ref{fluid}), but with hatted quantities replacing everywhere the original parameters. Finally, for simplicity we set the elementary  fermion charge and the UV $AdS_4$ length to one, 
\be
q_f = 1, \qquad L=1. 
\ee

With the ansatz (\ref{ans1}) and these field  and parameters redefinitions,  the field equations reduce to the system (\ref{eq:system}). In those equations, and in the rest of the article,  we have removed the hats from the rescaled fields to simplify the notation.

\section{First law and equation of state}\label{AppOnshell}

We will show the validity of the first law of thermodynamics for the backgrounds under consideration. 
We need to compute the on-shell action, that gives the free energy of the theory, the energy and the charge. 

The on-shell action can be shown, using the equations of motion (\ref{eq:system}), to reduce to a total derivative: 
\begin{equation}
\sqrt{-g} \, {\cal L}_{on-shell} = \partial_r \left( \frac{2 h h' - f'}{2 \sqrt{r^4 f g}} \right) \,.
\end{equation}
The renormalized action is the on-shell action supplemented by the Gibbons-Hawking and counterterms 
\begin{equation} 
\sqrt{-g}\,  {\cal L}_{on-shell} - {\cal L}_{GH} - 2 {\cal L}_{ct} \,,
\end{equation} 
with 
\begin{equation}
\begin{aligned} 
& {\cal L}_{GH} = \sqrt{- \gamma} \, K =  \frac{f'}{2 \sqrt{r^4 f g}} - \frac{2}{r^3} \sqrt{\frac{f}{g}} \,,\\
& {\cal L}_{ct} = \sqrt{- \gamma}  = \frac{\sqrt{f}}{r^2} \,.
\end{aligned}
\end{equation} 
where $K$ is the extrinsic curvature and $\gamma$ the induced metric on the hypersurface of constant $r$. 
One can easily check, plugging in the asymptotics in (\ref{uvasym}) that the on-shell action reduces to 
$M - \mu Q$. \\
The renormalized boundary stress-energy tensor can similarly be computed: 
\begin{subequations}
\begin{eqnarray} 
T_{a}^{b} &=& \sqrt{-\gamma} \left(K_a^b - (K+2) \delta_a^b\right)\,, \\
T_0^0 &=& \frac{\sqrt{f}}{r^2} \left(\frac{2 \sqrt{f}}{r^3 \sqrt{g}} -2\right) \,, \\
T_i^i &=&  \frac{\sqrt{f}}{r^2} \left(\frac{1}{r \sqrt{g}} - \frac{f'}{2 f \sqrt{g}} - 2\right) \,,
\end{eqnarray}
\end{subequations}
and one can check that $T^{00} = M$, so that the parameter $M$ in the solution can be identified with the energy, and $T_{a}^{a} = 0$, so the equation of state is that of a conformal theory.  

\section{Probe scalar field on electron star background} \label{probe}

In this Appendix we study the condensation of the charged scalar field of negative mass squared as a probe field on the background
of the electron star model at zero temperature.

\subsection{IR asymptotics}

The electron star solution in the IR has Lifshitz geometry
\begin{equation}
f(r) = \frac{1}{r^{2z}} \ , \ \ \ \ g(r) = \frac{g_\infty}{r^2} \ , \ \ \ \  h(r)
= \frac{h_\infty}{r^z} \,,
\end{equation}
where
\begin{equation}
h_\infty^2=(z-1)/z \ , \ \ \ g_\infty^2 = \frac{36(z-1)z^4}{[(1-m_f^2)z-1]^3\beta^2} \ .
\end{equation}
The asymptotic solution for $\psi$ on this background is
\begin{equation}
\psi \sim A_{-} \, r^{\Delta_\mathrm{IR}} + A_{+} \,
r^{(z+2)-\Delta_\mathrm{IR}}\ , \hskip0.3cm r\to\infty \,,
\end{equation}
where
\begin{equation}
\label{eq:delta-IR}
\Delta_\mathrm{IR} = \frac{1}{2}\left[(z+2)-
\sqrt{(z+2)^2-4g_\infty(h_\infty^2q^2-m_s^2)}\right] \ .
\end{equation}
At infinity, the second term is dominant.
\subsection{UV asymptotics}

Outside the star, the solution is RN-$AdS$ black hole. Asymptotically close to the UV
boundary, $\psi$ behaves like
\begin{equation}
\psi \sim B_- \, r^{3-\Delta_\mathrm{UV}} + B_+ \, r^{\Delta_\mathrm{UV}} \ , \hskip0.3cm r\to0 \,,
\end{equation}
where
\begin{equation}
\Delta_\mathrm{UV} = \frac{3}{2} + \sqrt{\frac{9}{4}+m_s^2} \ .
\end{equation}
%

\subsection{Condensation}

We look for solutions for which the scalar field is above the BF bound in the UV but below
the BF bound in the IR so that the IR solution leads to an instability and the condensation of the scalar field. Condensation occurs if

\begin{equation}
\begin{aligned}
\label{eq:UV-BF-bound}
-\frac{9}{4} < m_s^2 &< 0 \ , \\
m_s^2 &< m_c^2 \ , \ \ \ \ m_c^2 \equiv -\frac{(z+2)^2}{4g_\infty}+h_\infty^2 q^2 \ .
\end{aligned}
\end{equation}
\\
At zero charge, $q=0$, since $z>1$ and $g_\infty>1$ this is equivalent to have the
following condition on IR parameters~\cite{Edalati:2011yv}
\begin{equation}
-\frac{9}{4} < m_s^2 < - \frac{(z+2)^2}{4g_\infty}  \ .
\end{equation}
\\
Let us look at the non-zero charge case.
The IR dimension~(\ref{eq:delta-IR}) can be rewritten as
\begin{equation}
 \Delta_\mathrm{IR} = \frac{1}{2}\left[(z+2)-2\sqrt{g_\infty}\sqrt{m_s^2-m_c^2}\right] \ .
\end{equation}
\begin{figure}
\centering
\includegraphics[width=0.5\textwidth]{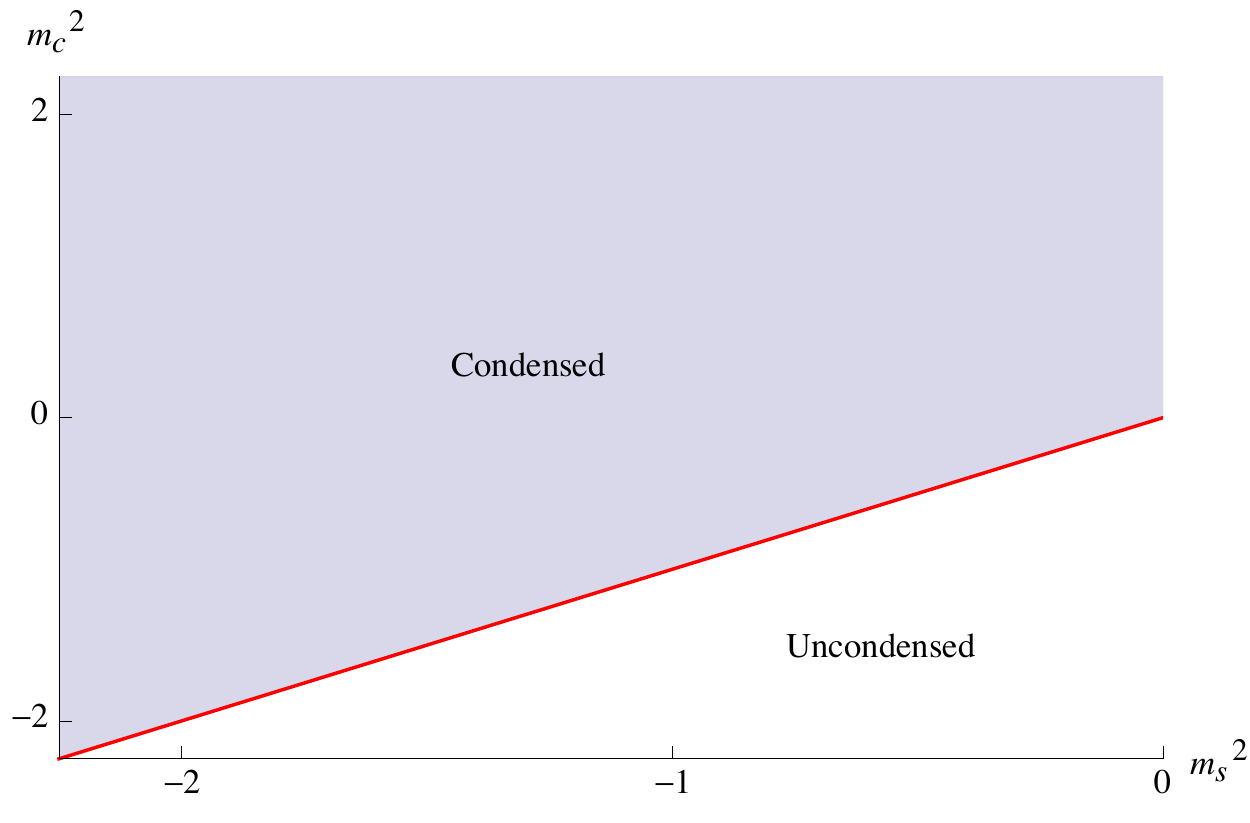}
\caption{Phase diagram for the probe scalar field at fixed electron star parameters $z$ and
$\beta$. The red line has equation $m_c^2=m_s^2$. For $m_c^2>0$, the scalar field condenses in
the IR for any $m_s^2<0$ above the BF bound. }
\label{fig:Phase-pert}
\end{figure}
\begin{figure}
\centering 
\includegraphics[width=0.5\textwidth]{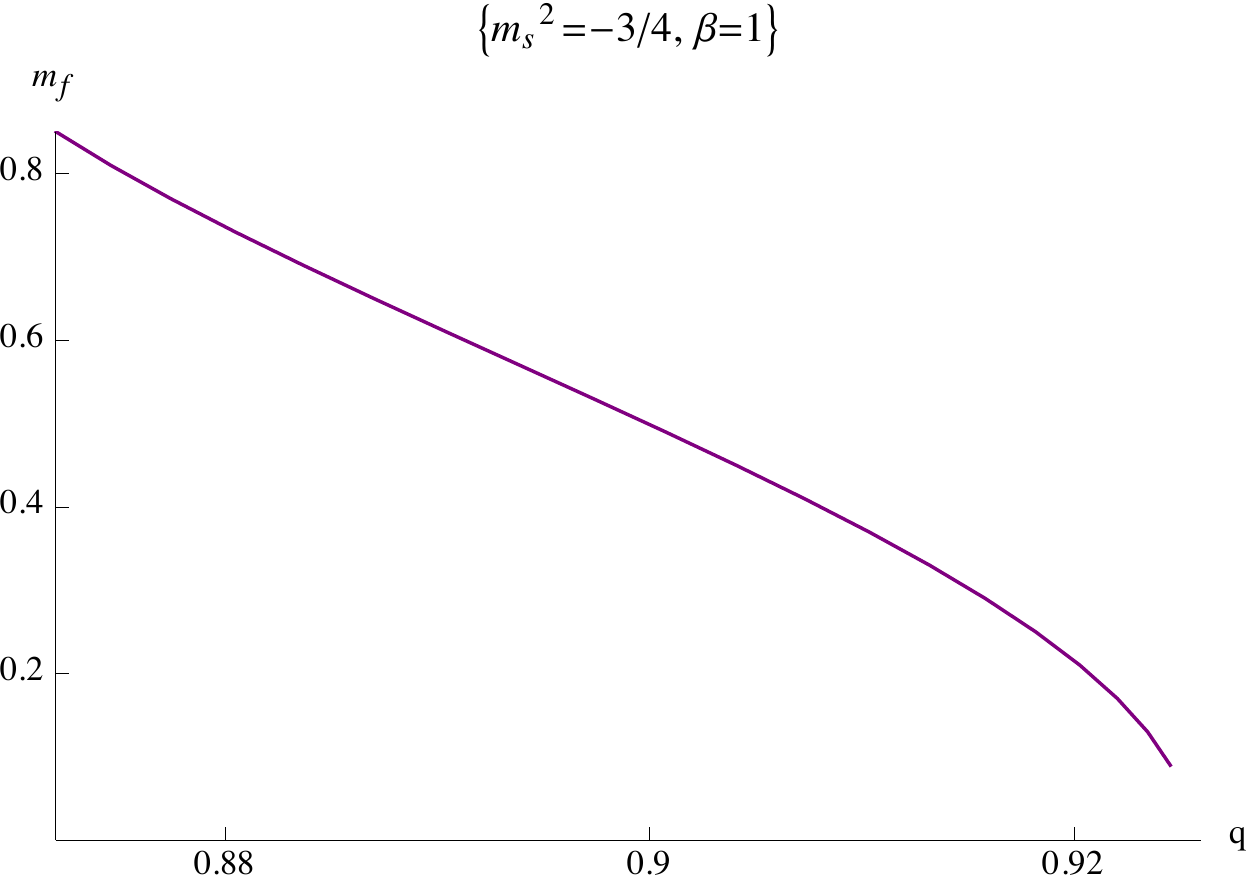}
\caption{The boundary of the stability region for the scalar in the ES background.}
\label{fig:ESinstability}
\end{figure}
The phase diagram for the condensation of the probe scalar field is given
in Fig.~\ref{fig:Phase-pert}. At fixed $m_s^2$ we can draw the phase boundary as a curve in the 
$(q,m_f)$ plane, as shown in Figure \ref{fig:ESinstability}. Comparing this to the backreacted phase diagram of 
Figure \ref{fig:phase} we notice that the two are not identical: when the probe approximation predicts the condensation, 
there is a CS solution, however there are also regions where the CS exists even though the ES is perturbatively stable, so the probe approximation 
can not reliably predict the phase boundary. 

\section{Irrelevant operators and IR Lifshitz solutions}\label{Lifshitz}

So far, we have investigated models in which the scalar field mass squared is negative, i.e. the dual boundary operator is a relevant one. Here, we briefly investigate the infrared geometry of the solutions  in the case $m_s^2 > 0$, in particular the interaction between the formation of the electron star and the condensation of the scalar field.  

We concentrate in particular on Lifshitz IR geometries, with the metric and gauge field coefficients of (\ref{sol}) of the form
\begin{equation}\label{lif1}
f = \frac{1}{r^{2z}} \ , \ \ \ \ g = \frac{g_0}{r^2} \ , \ \ \ \ h = \frac{h_0}{r^z}, \qquad r\to \infty, 
\end{equation}
where $f_0,g_0,h_0$ are positive constants and $z>2$,  and with asymptotically constant values of the scalar condensate and of the fluid charged density, 
\be\label{lif2}
\psi(r) \sim \psi_0, \qquad \sigma(r)\sim \sigma_0, \qquad r\to \infty . 
\ee

These geometries arise both from  $m_s^2>0$ charged scalar in the absence of the fluid, and as the IR geometry of the electron star with trivial condensate. In addition, we have found solutions with both fluid and scalar condensate turned on.  It is useful to rescale the scalar field by
\begin{equation}\label{lif3}
\psi \to \tilde{\psi}=q \psi \ .
\end{equation}
After this rescaling, the ansatz (\ref{lif1},\ref{lif2}) is a solution of the system (\ref{eq:system}) provided the constants satisfy the system: 
\begin{subequations}
\begin{align}
h_0 &= \sqrt{\frac{z-1}{z}} \ , \label{lif4a}\\
g_0 &= \frac{2z\sqrt{z-1}}{\sqrt{z}\sigma_0+\sqrt{z-1}\tilde\psi_0^2} \ , \label{lif4b} \\
\tilde{\psi}_0 &= \left(\frac{z}{z-1}\right)^{1/4} \left|
\frac{4\sqrt{z}\sqrt{z-1}(3+p_0)-(z+1)(z+2)\sigma_0}{2z\frac{m^2_s}{q^2}+ (4+z+z^2)}
\right|^{1/2} \ . \label{lif4c}
\end{align}
\end{subequations}
together with the constraint
\begin{equation}
\label{eq:mqpsi-constraint}
(m^2_s-q^2\mu_0^2)\tilde{\psi}_0 = 0 \ .
\end{equation}

Notice that after the rescaling (\ref{lif3}) the system  depends only on the  ratio $m_s/q$, and not on $m_s$ or $q$ independently.

There are three different possibilities:

\begin{enumerate} 
\item {\bf Star only:} If the scalar field does not condense, Einstein-Maxwell equations impose the constraint
\begin{equation}
4\sqrt{z}\sqrt{z-1}(3+p_0)-(z+1)(z+2)\sigma_0 = 0 \ , \qquad \tilde \psi_0 = 0 \,,
\end{equation}
which gives a non-trivial relation between $z$ and the fluid parameters $m_f$ and $\beta$.
This is  the zero temperature electron star solution~\cite{Hartnoll:2010gu}. 

\item {\bf Condensate only:}  If the scalar field condenses,  the
 constraint~(\ref{eq:mqpsi-constraint}) implies
that $m^2_s>0$ since $q^2\mu_0^2>0$, and the local chemical potential and dynamical exponents are
\begin{equation}
\mu_0 = \frac{m_s}{|q|} \ , \qquad   z = \frac{1}{1-m_s^2/q^2}. 
\end{equation}
When there is no star, i.e. $\sigma_0=p_0=0$, the condensate is given by
\begin{equation}
 \tilde{\psi}_0 = \frac{2\sqrt{3z}}{\sqrt{(z+1)(z+2)}} \ .
\end{equation}
This is the solution found in~\cite{Horowitz:2009ij}.

\begin{figure}[t]
\centering
\includegraphics[width=0.5\textwidth]{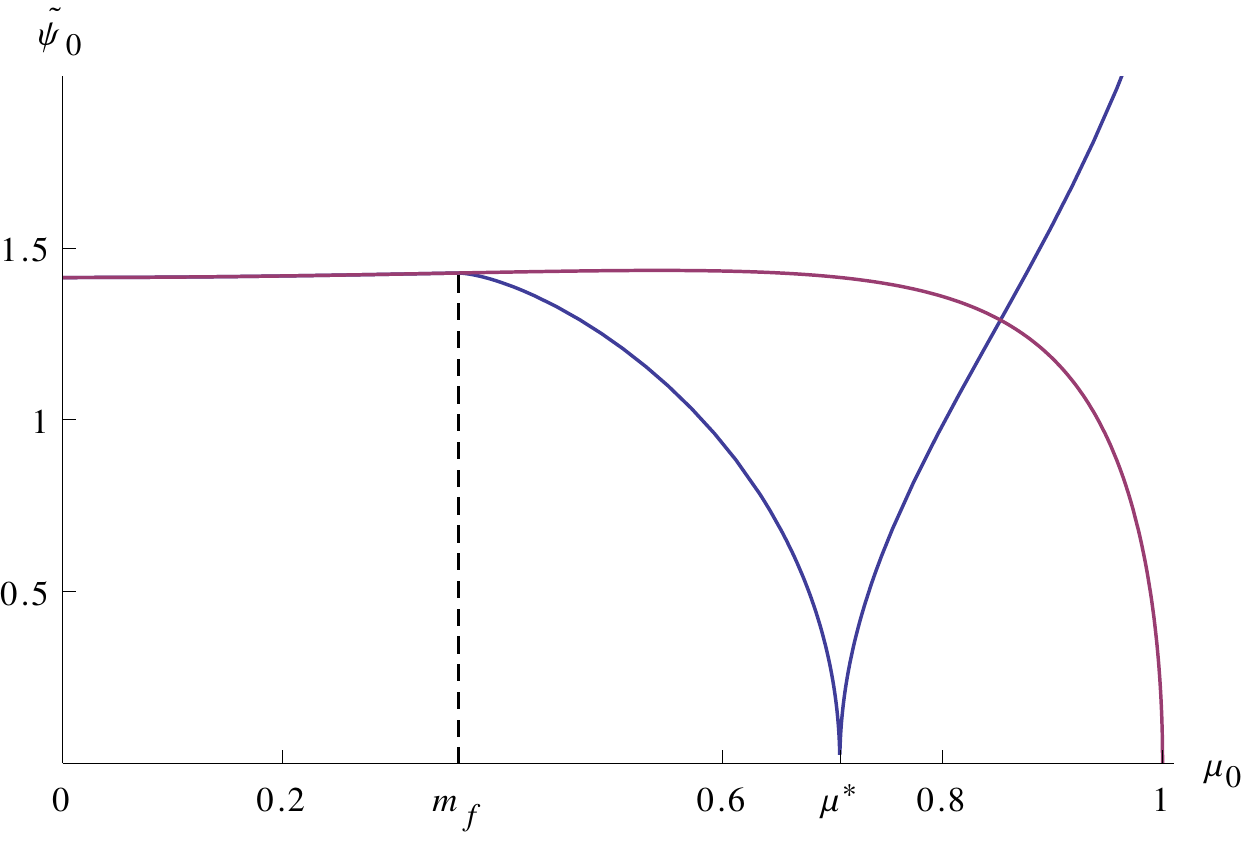}
\caption{The condensate $\tilde{\psi}_0$ as a function of $\mu_0=m_s/|q|$ for the Lifshitz solution
at $m_f=0.36$ and $\beta=20$. The blue and red lines represent the condensate with and without
the star respectively. When there is no star forming, for $\mu_0=1$, which
corresponds to $z=\infty$, the condensate vanishes while it tends to a constant
for $\mu_0=0$ ($z\to1$). When the star forms for $\mu_0>m_f$, there is a particular point
$\mu_0=\mu^*$ where the condensate vanishes. At this point, we recover the electron star
solution.
}
\label{fig:Lif-condensation}
\end{figure}

\item {\bf Coexistence phase:} The star and the condensate can coexist.
In this case, eq. (\ref{eq:mqpsi-constraint}) for $\tilde\psi_0\neq 0$ implies that the chemical potential is  again $\mu_0=m_s/|q|$, related to the dynamical exponent by (\ref{lif4a}). The precise value of  the condensate is given in terms of $m_f$, $\beta$ and $m_s/q$  by eq. (\ref{lif4c}).

\end{enumerate}

The space of solutions is depicted in Fig.~\ref{fig:Lif-condensation}, 
where we display  
 $\tilde{\psi}_0$ as a function of $\mu_0$ for $m_f=0.36$ and $\beta=20$. Notice that  for $\mu_0< m_f$ the star cannot form, and only the solution with the condensate alone exists. At $\mu_0 = m_f$ the star can start forming, so we have two branches. The special point $\mu_0 = \mu^*$,  which coincides with the vanishing of the right hand side of (\ref{lif4c}), admits both a pure  star and a pure condensate with the same Lifshitz exponent. 

Although the phase structure of these solutions is interesting, one does not expect that these  IR Lifshitz solutions will connect to the UV   asymptotically $AdS_4$ space: the squared mass
of the scalar field is positive, and   the operator dual to the scalar
field is an irrelevant operator in the UV. Thus, it is unlikely that one could find an RG flow from a UV $AdS$ region to these IR solutions.   


\bibliographystyle{JHEP}
\bibliography{myrefs}

\providecommand{\href}[2]{#2}\begingroup\raggedright\begin{thebibliography}{10}

\bibitem{Hartnoll:2009sz}
S.~A. Hartnoll, {\it {Lectures on holographic methods for condensed matter
  physics}},  {\em Class.Quant.Grav.} {\bf 26} (2009) 224002,
  [\href{http://xxx.lanl.gov/abs/0903.3246}{{\tt arXiv:0903.3246}}].

\bibitem{Herzog:2009xv}
C.~P. Herzog, {\it {Lectures on Holographic Superfluidity and
  Superconductivity}},  {\em J.Phys.} {\bf A42} (2009) 343001,
  [\href{http://xxx.lanl.gov/abs/0904.1975}{{\tt arXiv:0904.1975}}].

\bibitem{McGreevy:2009xe}
J.~McGreevy, {\it {Holographic duality with a view toward many-body physics}},
  {\em Adv.High Energy Phys.} {\bf 2010} (2010) 723105,
  [\href{http://xxx.lanl.gov/abs/0909.0518}{{\tt arXiv:0909.0518}}].

\bibitem{Hartnoll:2009ns}
S.~A. Hartnoll, J.~Polchinski, E.~Silverstein, and D.~Tong, {\it {Towards
  strange metallic holography}},  {\em JHEP} {\bf 1004} (2010) 120,
  [\href{http://xxx.lanl.gov/abs/0912.1061}{{\tt arXiv:0912.1061}}].

\bibitem{Hartnoll:2010gu}
S.~A. Hartnoll and A.~Tavanfar, {\it {Electron stars for holographic metallic
  criticality}},  {\em Phys.Rev.} {\bf D83} (2011) 046003,
  [\href{http://xxx.lanl.gov/abs/1008.2828}{{\tt arXiv:1008.2828}}].

\bibitem{Hartnoll:2011pp}
S.~A. Hartnoll and L.~Huijse, {\it {Fractionalization of holographic Fermi
  surfaces}},  {\em Class.Quant.Grav.} {\bf 29} (2012) 194001,
  [\href{http://xxx.lanl.gov/abs/1111.2606}{{\tt arXiv:1111.2606}}].

\bibitem{Adam:2012mw}
A.~Adam, B.~Crampton, J.~Sonner, and B.~Withers, {\it {Bosonic
  Fractionalisation Transitions}},  {\em JHEP} {\bf 1301} (2013) 127,
  [\href{http://xxx.lanl.gov/abs/1208.3199}{{\tt arXiv:1208.3199}}].

\bibitem{Hartnoll:2008vx}
S.~A. Hartnoll, C.~P. Herzog, and G.~T. Horowitz, {\it {Building a Holographic
  Superconductor}},  {\em Phys.Rev.Lett.} {\bf 101} (2008) 031601,
  [\href{http://xxx.lanl.gov/abs/0803.3295}{{\tt arXiv:0803.3295}}].

\bibitem{Hartnoll:2008kx}
S.~A. Hartnoll, C.~P. Herzog, and G.~T. Horowitz, {\it {Holographic
  Superconductors}},  {\em JHEP} {\bf 0812} (2008) 015,
  [\href{http://xxx.lanl.gov/abs/0810.1563}{{\tt arXiv:0810.1563}}].

\bibitem{Gubser:2008pf}
S.~S. Gubser and A.~Nellore, {\it {Low-temperature behavior of the Abelian
  Higgs model in anti-de Sitter space}},  {\em JHEP} {\bf 0904} (2009) 008,
  [\href{http://xxx.lanl.gov/abs/0810.4554}{{\tt arXiv:0810.4554}}].

\bibitem{Gubser:2009cg}
S.~S. Gubser and A.~Nellore, {\it {Ground states of holographics}},  {\em
  Phys.Rev.} {\bf D80} (2009) 105007,
  [\href{http://xxx.lanl.gov/abs/0908.1972}{{\tt arXiv:0908.1972}}].

\bibitem{Horowitz:2009ij}
G.~T. Horowitz and M.~M. Roberts, {\it {Zero Temperature Limit of Holographic
  Superconductors}},  {\em JHEP} {\bf 0911} (2009) 015,
  [\href{http://xxx.lanl.gov/abs/0908.3677}{{\tt arXiv:0908.3677}}].

\bibitem{Hartnoll:2010xj}
S.~A. Hartnoll, D.~M. Hofman, and A.~Tavanfar, {\it {Holographically smeared
  Fermi surface: Quantum oscillations and Luttinger count in electron stars}},
  {\em Europhys.Lett.} {\bf 95} (2011) 31002,
  [\href{http://xxx.lanl.gov/abs/1011.2502}{{\tt arXiv:1011.2502}}].

\bibitem{Hartnoll:2011dm}
S.~A. Hartnoll, D.~M. Hofman, and D.~Vegh, {\it {Stellar spectroscopy: Fermions
  and holographic Lifshitz criticality}},  {\em JHEP} {\bf 1108} (2011) 096,
  [\href{http://xxx.lanl.gov/abs/1105.3197}{{\tt arXiv:1105.3197}}].

\bibitem{Sachdev:2011ze}
S.~Sachdev, {\it {A model of a Fermi liquid using gauge-gravity duality}},
  {\em Phys.Rev.} {\bf D84} (2011) 066009,
  [\href{http://xxx.lanl.gov/abs/1107.5321}{{\tt arXiv:1107.5321}}].

\bibitem{Allais:2012ye}
A.~Allais, J.~McGreevy, and S.~J. Suh, {\it {A quantum electron star}},  {\em
  Phys.Rev.Lett.} {\bf 108} (2012) 231602,
  [\href{http://xxx.lanl.gov/abs/1202.5308}{{\tt arXiv:1202.5308}}].

\bibitem{Allais:2013lha}
A.~Allais and J.~McGreevy, {\it {How to construct a gravitating quantum
  electron star}},  \href{http://xxx.lanl.gov/abs/1306.6075}{{\tt
  arXiv:1306.6075}}.

\bibitem{Huijse:2011hp}
L.~Huijse and S.~Sachdev, {\it {Fermi surfaces and gauge-gravity duality}},
  {\em Phys.Rev.} {\bf D84} (2011) 026001,
  [\href{http://xxx.lanl.gov/abs/1104.5022}{{\tt arXiv:1104.5022}}].

\bibitem{Edalati:2011yv}
M.~Edalati, K.~W. Lo, and P.~W. Phillips, {\it {Neutral Order Parameters in
  Metallic Criticality in d=2+1 from a Hairy Electron Star}},  {\em Phys.Rev.}
  {\bf D84} (2011) 066007, [\href{http://xxx.lanl.gov/abs/1106.3139}{{\tt
  arXiv:1106.3139}}].

\bibitem{Liu:2013yaa}
Y.~Liu, K.~Schalm, Y.-W. Sun, and J.~Zaanen, {\it {Bose-Fermi competition in
  holographic metals}},  \href{http://xxx.lanl.gov/abs/1307.4572}{{\tt
  arXiv:1307.4572}}.

\bibitem{Gouteraux:2012yr}
B.~Gout\'eraux and E.~Kiritsis, {\it {Quantum critical lines in holographic
  phases with (un)broken symmetry}},  {\em JHEP} {\bf 1304} (2013) 053,
  [\href{http://xxx.lanl.gov/abs/1212.2625}{{\tt arXiv:1212.2625}}].

\bibitem{Hartnoll:2010ik}
S.~A. Hartnoll and P.~Petrov, {\it {Electron star birth: A continuous phase
  transition at nonzero density}},  {\em Phys.Rev.Lett.} {\bf 106} (2011)
  121601, [\href{http://xxx.lanl.gov/abs/1011.6469}{{\tt arXiv:1011.6469}}].

\bibitem{Sonner:2009fk}
J.~Sonner, {\it {A Rotating Holographic Superconductor}},  {\em Phys.Rev.} {\bf
  D80} (2009) 084031, [\href{http://xxx.lanl.gov/abs/0903.0627}{{\tt
  arXiv:0903.0627}}].

\end{thebibliography}\endgroup

\end{document}